\documentclass[letterpaper,12pt]{article}
\pdfoutput=1

\usepackage{amsmath}
\usepackage{amsfonts}
\usepackage{amssymb}
\usepackage{graphicx}
\usepackage{color}
\usepackage{slashed}
\usepackage{dsfont}

\usepackage{graphicx}
\usepackage{epstopdf}
\usepackage[body={6.5in, 9in}, right=1in, top=1in]{geometry}
\usepackage{amssymb}
\usepackage{amsmath} 
\usepackage[numbers,sort&compress]{natbib}

\numberwithin{equation}{section}

\newcommand\ignore[1]{}
\def\one{{\,\hbox{1\kern-.8mm l}}}

\def\({\left(}
\def\){\right)}

\newcommand{\rmd}{\mathrm{d}}
\newcommand{\rd}{{\rm d}}

\newcommand{\Cset}{{\,\,{{{^{_{\pmb{\mid}}}}\kern-.45em{\mathrm C}}}}}

\newcommand{\be}{\begin{equation}}
\newcommand{\bea}{\begin{eqnarray}}

\newcommand{\ee}{\end{equation}}
\newcommand{\eea}{\end{eqnarray}}

\newcommand{\vp}{\varphi}
\newcommand{\vphi}{\varphi}

\providecommand{\lsim}{\lesssim}

\newcommand{\Comment}[1]{{}}
\definecolor{MyDarkBlue}{rgb}{0.15,0.15,0.45}
\usepackage[linktocpage=true]{hyperref}
\hypersetup{
colorlinks=true,
citecolor=MyDarkBlue,
linkcolor=MyDarkBlue,
urlcolor=MyDarkBlue,
}

\begin{document}
\def\thefootnote{\fnsymbol{footnote}}

\begin{center}
\Large{\textbf{Consistency Relations for the Conformal Mechanism}} \\[0.5cm]
 
\large{Paolo Creminelli$^{\rm a}$, Austin Joyce$^{\rm b}$, Justin Khoury$^{\rm b}$, and Marko Simonovi\'c${}^{\rm c,d}$}
\\[0.5cm]

\small{
\textit{$^{\rm a}$ Abdus Salam International Centre for Theoretical Physics\\ Strada Costiera 11, 34151, Trieste, Italy}}

\vspace{.2cm}

\small{
\textit{$^{\rm b}$ Center for Particle Cosmology, Department of Physics and Astronomy, \\ University of Pennsylvania, Philadelphia, PA 19104, USA}}

\vspace{.2cm}

\small{
\textit{$^{\rm c}$ SISSA, via Bonomea 265, 34136, Trieste, Italy}}

\vspace{.2cm}

\small{
\textit{$^{\rm d}$ Istituto Nazionale di Fisica Nucleare, Sezione di Trieste, I-34136, Trieste, Italy}}

\end{center}

\vspace{.8cm}

\hrule \vspace{0.3cm}
\noindent \small{\textbf{Abstract}\\
We systematically derive the consistency relations associated to the non-linearly realized symmetries of theories with spontaneously broken conformal symmetry but with a linearly-realized de Sitter subalgebra. These identities relate $(N+1)$-point correlation functions with a soft external Goldstone to $N$-point functions. These relations have direct implications for the recently proposed conformal mechanism for generating density perturbations in the early universe. We study the observational consequences, in particular a novel one-loop contribution to the four-point function, relevant for the stochastic scale-dependent bias and CMB $\mu$-distortion.}
\vspace{0.3cm}
\noindent
\hrule
\def\thefootnote{\arabic{footnote}}
\setcounter{footnote}{0}


\section{Introduction}

Although non-linearly realized symmetries have long been known to be a powerful tool in many areas of physics, only recently has this technology been brought to bear on theories of the early universe. For example, it has been realized that perturbations in single-field inflation can be described most generally by the effective field theory of spontaneously broken time diffeomorphisms~\cite{Cheung:2007st,Creminelli:2006xe}.
Single-field inflation can also be understood in terms of {\it global} symmetries as the spontaneous breaking of the $SO(4,1)$ conformal symmetry of $\mathds{R}^3$ down to spatial translations and rotations~\cite{Creminelli:2012ed,Hinterbichler:2012nm}. The corresponding Goldstone field is $\zeta$, the curvature perturbation of uniform-density hypersurfaces. Moreover, the well-known consistency relations~\cite{Maldacena:2002vr,Creminelli:2004yq,Cheung:2007sv,Senatore:2012wy, Creminelli:2011rh}, which constrain the soft limit of correlation functions, arise as Ward identities for the non-linearly realized symmetries~\cite{Assassi:2012zq,LamKurt,Goldberger:2013rsa}. Additionally, symmetry considerations have proven to be a powerful tool in analyzing correlation functions of spectator fields in inflation: both gravitons~\cite{Maldacena:2011nz} and scalar field perturbations~\cite{Antoniadis:1996dj,Antoniadis:2011ib,Creminelli:2011mw,Kehagias:2012pd,Kehagias:2012td} are constrained to have conformally-invariant correlators at late times.

Recently, it has been realized that the de Sitter symmetries --- including dilation invariance, responsible for the scale invariance of perturbations --- need not correspond to the isometries of a physical metric, as in inflation, but instead can arise as the unbroken sub-algebra of spontaneously broken conformal symmetry where the dynamical metric is nearly flat~\cite{Rubakov:2009np,Creminelli:2010ba,Hinterbichler:2011qk}. In this {\it conformal mechanism}, some fields in the conformal field theory (CFT) acquire specific time-dependent expectation values, which breaks the $SO(4,2)$ conformal symmetry on (approximate) Minkowski space down to its de Sitter subgroup:
\be
\label{symmbreak}
SO(4,2) \longrightarrow SO(4,1) \,.
\ee
As a result of couplings dictated by conformal invariance, spectator fields in the theory evolve in a fictitious de Sitter background, and consequently massless fields
acquire a nearly scale invariant spectrum of perturbations.\footnote{It is well-known that weight-zero fields are forbidden by unitary bounds~\cite{Mack:1975je}, but these assume a stable conformally invariant vacuum. The conformal mechanism does not assume such a vacuum, only a time-dependent symmetry-breaking background. The conformal vacuum can be unstable or not exist.} The de Sitter expansion is fictitious; the physical, Einstein-frame metric describes a universe which is very slowly contracting or expanding. Incidentally, such slow evolution drives the universe to be increasingly flat, homogeneous and isotropic~\cite{Gratton:2003pe}, thereby addressing the well-known problems of standard big bang cosmology. A robust prediction of the scenario is the absence of gravitational waves --- because the universe is approximately static, tensor modes are not appreciably excited. As with any mechanism relying on multiple fields, a generic prediction is a significant level of non-Gaussianity in the squeezed/local limit. 

An example of this mechanism is Galilean Genesis~\cite{Creminelli:2010ba}. This scenario is based on the conformal galileons~\cite{Nicolis:2008in}, a class of conformally-invariant, higher-derivative scalar field theories, which nevertheless only propagate one degree of freedom. Galilean Genesis capitalizes on the fact that galileons can violate the Null Energy Condition (NEC)~\cite{Nicolis:2009qm} in a stable manner, to describe a universe which expands from an asymptotically flat initial state. One drawback of the original scenario is that fluctuations around (slight deformations) of the NEC-violating background propagate superluminally. Superluminality can be avoided through explicit breaking of the conformal group, preserving only dilation~\cite{Creminelli:2012my}. Alternatively, the Dirac--Born--Infeld (DBI) generalization of the scenario~\cite{Hinterbichler:2012fr,Hinterbichler:2012yn} preserves the full conformal symmetries, but, thanks to a different non-linear realization of the conformal algebra, features perturbations that propagate strictly sub-luminally around the NEC-violating background. Another incarnation of the conformal mechanism is the $U(1)$-invariant model~\cite{Craps:2007ch,Rubakov:2009np,Hinterbichler:2011qk}, which describes a complex scalar field rolling down a negative quartic potential. The resulting cosmology is a phase of slow contraction before a NEC-violating phase that leads to the big bang.

In fact, the conformal mechanism is more general than these particular incarnations. All the important physics --- in particular scale invariance --- is fixed by the symmetry breaking pattern~(\ref{symmbreak}),
irrespective of the microphysical details. Applying the technology of the coset construction~\cite{Coleman:1969sm,Callan:1969sn} for space-time symmetries \cite{volkov}, the most general effective action describing the Goldstone field $\pi$ and other  ``matter" fields, including weight-zero fields, was derived in~\cite{Hinterbichler:2012mv} systematically in powers of derivatives. By construction, the effective action linearly realizes the $SO(4,1)$ de Sitter algebra\footnote{We will be somewhat loose about notation and refer to both Lie groups and their corresponding algebras in the same way, {\it e.g.}, $SO(4,2)$ for the conformal group/algebra.} and non-linearly realizes the $SO(4,2)$ algebra. Many key properties of the scenario follow immediately from this effective action, {\it e.g.}, the scale invariance of massless fluctuations, and the fact that the time-dependent background is a dynamical attractor. 

In this paper we focus on the implications of the symmetry breaking pattern~(\ref{symmbreak}) on various correlation functions. As the models under consideration enjoy linearly-realized de Sitter symmetries, we begin in Sec.~\ref{desittersymms} by reviewing some elementary properties of field theory on de Sitter space, focusing in particular on the late-time action of the de Sitter isometries and their corresponding constraints on correlation functions.
\begin{figure}
\centering
\includegraphics[width=5.5in]{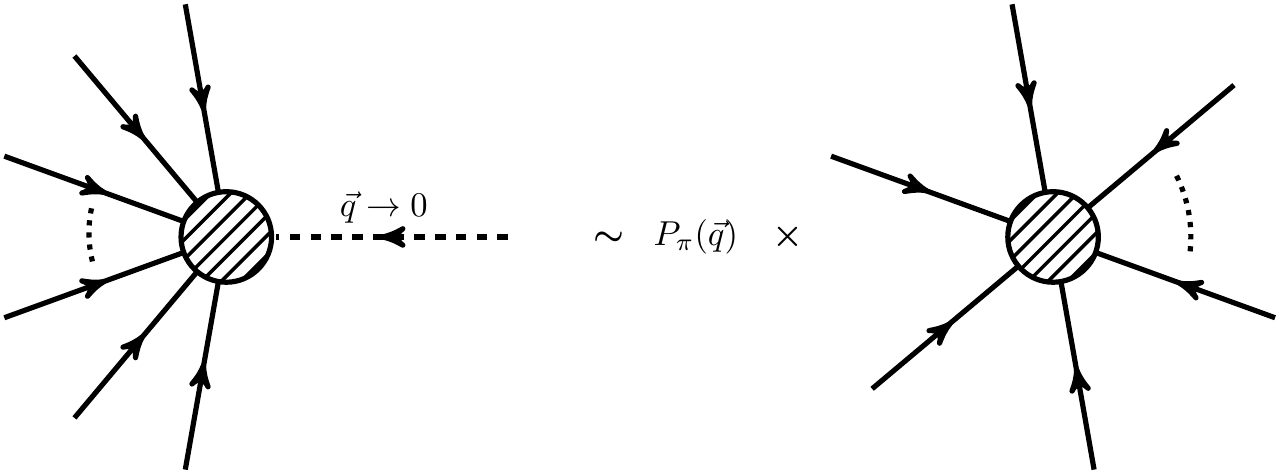}
\caption{\label{fig:consistency} \small Schematic representation of the consistency relations: $(N+1)$-point functions with a soft external $\pi$ leg are related to $N$-point functions.}
\end{figure}
%

Our primary interest, however, lies in understanding how the non-linearly realized symmetries --- time translation $P^0$, boosts $J^{0i}$, and the temporal component of the special conformal transformations $K^0$ --- act on correlation functions.  (Notice that the number of broken generators does not correspond to the number of Goldstones---only one in the case at hand---as we are dealing with a space-time symmetry \cite{Low:2001bw}.) These symmetries are discussed in Sec. \ref{nonlinearso42symms}. Analogously to single-field inflation, we will find that these result in {\it consistency relations} relating $(N+1)$-point correlation functions with a soft external $\pi$ to a (broken) symmetry transformation on the $N$-point functions without the soft $\pi$. In Sec.~\ref{derivationoftheconsistencyrelation}, we derive the equal-time master consistency relation, which is the main result of the paper (see Fig.~\ref{fig:consistency}):
\be
\label{introequation}
\langle \pi_{\vec q}\phi_{\vec k_1}\ldots \phi_{\vec k_N} \rangle'_{q\rightarrow 0} \overset{\rm avg}{=} -P_\pi(q)\(1 + \frac{1}{N} q_i \sum_a \partial_{k_{ai}}  + \frac{1}{6N}\vec q^{\;2} \sum_a \partial_{k_a}^2 \)  t \frac{\rmd}{\rmd t} \langle \phi_{\vec k_1}\ldots \phi_{\vec k_N} \rangle' \;,
\ee
where the $\phi$'s are fields of arbitrary mass, the $'$ indicates a correlation function without the momentum-conserving delta function, $P_\pi$ is the $\pi$ power spectrum and $\overset{\rm avg}{=}$ indicates equality after performing an angular average only over the terms quadratic in $q$.
Since there is only one Goldstone for 5 broken symmetries, the above consistency relation constrains different powers of $q$ in the soft limit $\vec{q}\rightarrow 0$.
The ${\cal O}(q^0)$ part of the constraint results from $P^0$, the ${\cal O}(q^1)$ part from $J^{0i}$, and the ${\cal O}(q^2)$ part from $K^0$. We derive~(\ref{introequation}) using the background-wave method~\cite{Creminelli:2004yq}, in which the long-wavelength $\pi$ mode is treated as a classical background for the short-wavelength modes; in a follow-up paper, we will show how~(\ref{introequation}) arises as a consequence of the Ward identities for the broken symmetries, analogous to the classic soft-pion theorems of chiral perturbation theory~\cite{Adler:1964um,Weinberg:1966kf}. We also check the validity of~(\ref{introequation}) in a myriad of cases using explicit correlation functions derived from the general effective action of~\cite{Hinterbichler:2012mv}.

The consistency relation~(\ref{introequation}) opens up the possibility of strong observational tests of the conformal mechanism, which is the focus of Sec.~\ref{softlinesrubakov}. In much the same way that observation of $f_{\rm NL}$ in the squeezed configuration would rule out all single-field models of the early universe, observation of a violation of one of these consistency relations would rule out the production of density perturbations by the conformal mechanism. An important caveat in making contact with observations is the observability of the $\pi$ mode --- although $\pi$ acquires a strongly red-tilted spectrum, its contribution to the adiabatic mode is actually strongly blue-tilted and hence is negligible on scales probed by cosmological observations. (The observed $\zeta$ is dominated by the scale invariant contribution from conversion~\cite{Wang:2012bq} of the weight-zero fields.) However, the consistency relation~(\ref{introequation}) does have observational implications for correlators in two ways. The first is for correlation functions with a soft internal line. In the limit where $M$ external momenta sum up to a soft total momentum, the amplitude is dominated by soft-$\pi$ exchange, thanks to its strongly red-tilted spectrum, and factorizes into a product of $(M + 1)$-point and $(N-M + 1)$-point functions, each with a soft $\pi$ insertion:
\be
\langle \phi_{\vec k_1}\ldots \phi_{\vec k_N} \rangle'_{q\rightarrow 0} =\frac{1}{P_\pi(q)} \langle \pi_{-\vec{q}} \phi_{\vec k_1}\ldots \phi_{\vec k_M} \rangle'_{q\rightarrow 0} \langle \pi_{\vec{q}} \phi_{\vec k_{M+1}}\ldots \phi_{\vec k_N} \rangle'_{q\rightarrow 0} \,.
\ee
The consistency relation~(\ref{introequation}) then constrains each soft amplitude in the product. We show how this constrains the form of the four-point function for massless spectator fields $\chi$ in the limit where the sum of two momenta is soft, reproducing the behavior explicitly calculated in \cite{Libanov:2010nk, Libanov:2010ci, Libanov:2011bk}. We also find a novel contribution to the four-point function of spectator fields from a one-loop exchange of two $\pi$'s. Despite being a one-loop effect, it is enhanced compared to the tree-level amplitude by inverse powers of $q$ so that it dominates in the soft limit. Its momentum dependence is of the $\tau_{\rm NL}$ form
\be
\tau_{\rm NL} \sim \frac{{\cal P}_\pi^2}{{\cal P}_\zeta} \log\frac{q}{\Lambda} \;,
\ee
(${\cal P}$ indicates the normalization of the power spectrum, {\em i.e.}~stripped of its momentum dependence)
so that we can expect signals in the stochasticity of the scale-dependent bias \cite{Baumann:2012bc} and the power spectrum of CMB $\mu$-distortion \cite{Pajer:2012vz}. Although $\pi$ itself is not observable, there is still an observable signature of soft external $\pi$ fields; as was pointed out in \cite{Libanov:2010nk, Libanov:2010ci, Libanov:2011bk}, the presence of a long $\pi$ mode leads to statistical anisotropy in spectator correlation functions, in particular it leads to anisotropy of the power spectrum of $\chi$:
\be
\langle \chi_{\vec k} \chi_{-\vec k} \rangle'_{\pi_{\vec q}}=\langle \chi_{\vec k} \chi_{-\vec k} \rangle'\left( 1 + c_1 \frac{\sqrt{{\cal P}_\pi}}{2\pi} \frac{H_0}{k}(3 \cos^2 \theta -1 ) + c_2 \frac{ 3 \,{\cal P}_\pi}{4 \pi^2} \cos^2\theta \log \frac{H_0}{\Lambda} \right) \;,
\ee
where $\theta$ is the angle between $\vec{k}$ and $\vec{q}$, $H_0$ is the present-day Hubble parameter, and $c_{1,2}$ are constant coefficients which depend on the particular realization of the long-wavelength $\pi$ modes. We will show that this anisotropy follows solely from symmetry considerations, and is therefore also a generic prediction of the conformal mechanism.

Since the scenario relies on de Sitter symmetries, it is natural to wonder whether the conformal symmetries can also be present in the inflationary context. The short answer is yes, but the inflationary realization is far less natural. Note first that the action we write down for $\pi$ cannot be the action for perturbations of the inflaton --- they have linearly-realized $SO(4,1)$ symmetry, and thus do not fit into the effective field theory of inflation framework. If $\pi$ were the perturbations of the inflaton, the corresponding de Sitter space would be eternal, there is no clock telling inflation when to end. We are therefore led to consider a situation where $\pi$ is merely a spectator field\footnote{Of course, there is another alternative, namely for the field $\pi$ itself to acquire a profile and act as the inflaton. While possible, it is easy to check that the potential for $\pi$ required by conformal invariance only supports inflation over a limited region in field space, $\pi \lsim M_{\rm Pl}$, corresponding to approximately 1 e-fold of inflation.} in multi-field inflation --- its presence therefore gives additional structure to the spectator sector and all of the same consistency relations will hold provided the entire spectator sector couples in a conformally invariant way to $\pi$. However, this is a somewhat artificial situation. We can of course impose this additional structure, but it does not buy us anything new --- whereas in the conformal mechanism this structure is a {\it necessary} and {\it natural} consequence of the mechanism.

Various subtleties of our calculations are addressed in detail in Sec.~\ref{extrastuff}, including complications introduced by the fact that $\pi$ is a field of negative 3d conformal weight in a de Sitter background,
the ambiguity (or lack thereof) of off-shell versus on-shell correlators in the consistency relation, and the translation from the consistency relation~(\ref{introequation}) to the form expected from the Ward identities
for spontaneously broken conformal symmetries. We summarize our results and discuss future spin-off directions in Sec.~\ref{conclu}. Various appendices collect results peripheral, but important, to our main line of development: Appendix~\ref{correlatoroperators} derives the action of the late-time de Sitter isometries on correlation functions, Appendix~\ref{constructionofactions} reviews the construction of actions for spontaneously broken conformal symmetry, and Appendix~\ref{correlationfncomputation} computes the correlation functions needed to adequately verify the consistency relation.

\section{Linearly realized SO(4,1) and 3d conformal transformations}
\label{desittersymms}
The scenario relies on linearly realized $SO(4,1)$ invariance, so we begin by reviewing some properties of scalar fields on de Sitter space. Throughout, we will work in the planar
slicing of de Sitter space, where the line element takes the form\footnote{Here we use $t$ as the conformal time coordinate on de Sitter space in order to emphasize the connection with models in which de Sitter arises as a fictitious background from broken conformal invariance.} 
\be
\rd s^2 = \frac{1}{H^2 t^2}\left(-\rd t^2+\rd\vec x^2\right)~,
\label{planedesitter}
\ee
where $t<0$ is conformal time. This is identical to the situation in multi-field inflation, where spectator fields feel a de Sitter geometry and do not back-react appreciably. A key difference, as emphasized in the Introduction, is that de Sitter space is a fake geometry in the conformal mechanism --- the actual, Einstein-frame metric is slowly evolving. Nevertheless, for the purpose of this discussion we can remain
agnostic as to whether or not the background de Sitter corresponds to the actual metric. 

The de Sitter metric~(\ref{planedesitter}) corresponds to a maximally symmetric space-time and therefore enjoys $10$ isometries. Six of these are the familiar translations and rotations of the flat spatial slices:
\begin{align}
x^i &\longrightarrow x^i+\alpha^i~;\\
x^i &\longrightarrow J^i_{~j}x^j~.
\label{transrot}
\end{align}
Additionally, de Sitter space is invariant under a dilation of both spatial and time coordinates
\be
x^\mu \longrightarrow \lambda x^\mu~.
\label{dil}
\ee
Finally, it is invariant under the simultaneous transformation of space and time as
\begin{align}
t &\longrightarrow t - 2t(\vec b\cdot\vec x)~;\\
x^i &\longrightarrow x^i+b^i(-t^2+\vec x^2)-2x^i(\vec b\cdot\vec x)~,
\label{sct}
\end{align}
where $b^i$ is a real-valued $3$-vector.

Next, consider a free scalar field on the de Sitter background:
\be
S = \int\rd^4x\sqrt{-g}\left(-\frac{1}{2}(\partial\phi)^2-\frac{m_\phi^2}{2}\phi^2\right)~.
\label{scalarondS}
\ee
The de Sitter isometries act on $\phi$ as follows:  spatial rotations and translations~(\ref{transrot}) act in the usual way,
\bea
\nonumber
\delta_{P_i}\phi &=& -\partial_i\phi\;; \\
\delta_{J_{ij}}\phi &=& (x_i\partial_j - x_j\partial_i)\phi~,
\label{transrotfield}
\eea
while the remaining four isometries~(\ref{dil}) and~(\ref{sct}) act as
\bea
\nonumber
\delta_{D}\phi &=& -(-t\partial_t+\vec x\cdot\vec\partial)\phi\;; \\
\delta_{K_i}\phi &=& -\left(-2x_it\partial_t+2x_i\vec x\cdot\vec\partial-(-t^2+x^2)\partial_i\right)\phi~.
\label{dilsct}
\eea
We are interested in how these transformations act at late times ($t \rightarrow 0$). In Fourier space, the equation of motion that follows
from the above action in the coordinates~\eqref{planedesitter} is 
\be
\ddot{\phi}_k -\frac{2}{t}\dot{\phi}_k +\left(k^2+\frac{m_\phi^2}{H^2t^2}\right)\phi_k = 0~,
\ee
with the well-known solution given by Hankel functions. In the long-wavelength ($k\rightarrow 0$) limit, the time dependence of the mode functions simplifies to
\be
\label{free_evolution}
\phi_k \sim t^{\Delta_\pm}~,~~~~~~~{\rm with}~~\Delta_\pm = \frac{3}{2}\pm\sqrt{\frac{9}{4}-\frac{m^2_\phi}{H^2}}~.
\ee
Assuming $m^2_\phi\leq 9H^2/4$, the growing mode corresponds to $\Delta_-\equiv\Delta$, and the time dependence of the field is $\phi\sim t^\Delta$ as $t \rightarrow 0$.
In this limit, we can therefore replace $t \partial_t \to\Delta$ in the transformation rules~(\ref{dilsct}) and neglect  ${\cal O}(t^2)$ terms to obtain
\bea
\nonumber
\delta_{D}\phi &=& \left(\Delta-\vec x\cdot\vec\partial\right)\phi\;; \\
\delta_{K_i}\phi &=& \left(2\Delta x_i-2x_i\vec x\cdot\vec\partial+x^2\partial_i\right)\phi~.
\label{dilsctlatetimes}
\eea
These are recognized respectively as spatial dilations and special conformal transformations for a field of conformal weight $\Delta$. Combined with the spatial rotations and translations~(\ref{transrotfield}), they form the conformal group on $\mathbb R^3$. Therefore, correlation functions of fields on de Sitter must be invariant under conformal transformations of Euclidean 3-space on the future boundary~\cite{Maldacena:2011nz, Creminelli:2011mw,Antoniadis:1996dj,Antoniadis:2011ib,Kehagias:2012pd}, which is of course the basis of the proposed dS/CFT correspondence~\cite{Strominger:2001pn}. As reviewed in Appendix~\ref{correlatoroperators}, these symmetries act on $N$-point correlation functions in Fourier space as
\begin{align}
\nonumber
\label{conformaltransdeltas}
\delta_D{\cal A}'_N &= \left(\sum_{a=1}^N \left(\Delta_a -3-\vec k_a\cdot\vec\partial_{k_a}\right)+3\right){\cal A}'_N = 0~;\\
\delta_{K^i}{\cal A}'_N &= i\sum_{a=1}^N\left(2(\Delta_a -3)\partial_{k_a^i}+k^i_a\vec\partial_{k_a}^2-2\vec k_a\cdot\vec\partial_{k_a}\partial_{k_a^i}\right){\cal A}'_N = 0~.
\end{align}
These constrain the form that the correlation functions take in momentum space. Here we assumed that the free evolution \eqref{free_evolution} dominates at late times. If this is not the case, as we discuss in Sec.~\ref{extrastuff}, one cannot trade the time dependence of correlation functions for $\Delta$'s.

\section{Non-linearly realized conformal symmetry}
\label{nonlinearso42symms}

Returning to the action~(\ref{scalarondS}), we will be led to consider a particular choice of mass, corresponding to the quadratic action
\be
S_\pi = M_\pi^2\int\rd^4x\sqrt{-g}\left(-\frac{1}{2}(\partial\pi)^2+2H^2\pi^2\right)~,
\label{goldstonequadraticaction}
\ee
which has $m_\pi^2 = -4H^2~(\Delta=-1)$. In fact, this action is the quadratic action for the Goldstone of broken conformal symmetry. We briefly review how \eqref{goldstonequadraticaction} can arise naturally on an {\it effective} de Sitter space through the spontaneous breaking of conformal symmetry in a flat space quantum field theory. We will consider two simple examples, and see that they both lead to the quadratic action~\eqref{goldstonequadraticaction} for fluctuations.

\begin{itemize}

\item {\bf Negative quartic potential}. As a first example, we consider $\phi^4$ field theory on flat space with `wrong-sign'  coupling constant~\cite{Rubakov:2009np,Hinterbichler:2011qk,Hinterbichler:2012mv,Craps:2007ch}:
\be
S = \int\rd^4x\left(-\frac{1}{2}(\partial\phi)^2 + \frac{\lambda}{4}\phi^4\right)~\,,
\label{negquarticaction}
\ee
where $\lambda > 0$. This action is classically invariant under conformal transformations,\footnote{Incidentally, this theory is also asymptotically free.} consisting of 
space-time translations $P^\mu$, Lorentz transformations $J^{\mu\nu}$, space-time dilation $D$, and special conformal transformations $K^\mu$. These 15 symmetries form the algebra
$SO(4,2)$. 

Assuming a homogeneous field profile, the equation of motion for $\phi$ reduces to $\ddot\phi - \lambda\phi^3 = 0$, with zero-energy solution
\be
\bar\phi(t) = \sqrt{\frac{2}{\lambda}}\frac{1}{(-t)} \equiv \frac{M_\pi}{H(-t)}~.
\label{1otsoln}
\ee
Here $H$ will soon be understood as playing the role of a fake Hubble constant. This background, which describes the field starting out from the top of the potential in the asymptotic past and rolling down subsequently, spontaneously breaks 5 of the original conformal symmetries, namely $P^0$, $J^{0i}$ and $K^0$. The 10 unbroken symmetries, $D$, $P^i$, $J^{ij}$, and $K^i$, form the $SO(4,1)$ de Sitter subalgebra. 
Expanding about the $1/t$ background as $\phi = \bar\phi +\vp$, we obtain the following quadratic action for fluctuations:
\be
S = \int\rd^4 x\left(-\frac{1}{2}(\partial\vp)^2+ \frac{3}{t^2}\vp^2\right)~.
\ee
Through a field redefinition, $\phi = \bar\phi+\vp = \bar\phi e^\pi = \frac{M_\pi}{H(-t)}e^\pi$, and introducing an effective de Sitter metric
\be
g_{\mu\nu}^{\rm eff} \equiv \frac{1}{H^2t^2}\eta_{\mu\nu}~,
\label{dSeff}
\ee
the quadratic action becomes
\be
S =M_\pi^2\int\rd^4x\sqrt{-g_{\rm eff} }\left(-\frac{1}{2}g_{\mu\nu}^{\rm eff} \partial^\mu\pi\partial^\nu\pi+2H^2\pi^2\right)~,
\label{piquad}
\ee
which is precisely of the form~\eqref{goldstonequadraticaction}. As advocated, this action arises through the spontaneous symmetry breaking $SO(4,2) \to SO(4,1)$.

\item {\bf Galilean Genesis}. In its simplest guise, Galilean Genesis~\cite{Creminelli:2010ba} is achieved with a (wrong-sign) kinetic term plus a cubic conformal galileon term:
\be
S = \int\rd^4 x\left( f^2 e^{2\Pi}(\partial\Pi)^2+\frac{f^3}{\Lambda^3}\square\Pi(\partial\Pi)^2+\frac{f^3}{2\Lambda^3}(\partial\Pi)^4\right)~,
\ee
where the scales $f, \Lambda$ have dimensions of mass, and the scalar field $\Pi$ is dimensionless. This action is also invariant under the conformal group $SO(4,2)$, but in this case dilations and special conformal transformations act non-linearly to start with. The equation of motion following from this action admits a background solution of the form \cite{Creminelli:2010ba}
\be
e^{\bar\Pi} = \frac{1}{H(-t)}~,~~~~~~{\rm where} ~~~H^2 \equiv \frac{2\Lambda^3}{3f}~.
\ee
This solution preserves the de Sitter subgroup of the conformal group. Perturbing about this solution $\Pi = \bar\Pi +\pi$, and introducing the effective de Sitter metric~(\ref{dSeff}),
the quadratic action takes exactly the form~(\ref{piquad}).

\end{itemize}

More generally, the action for $\pi$ follows solely from the pattern of symmetry breaking. In \cite{Hinterbichler:2012mv}, the most general action for the goldstone of the breaking pattern $SO(4,2)\to SO(4,1)$ was constructed. To cubic order in the fields and second order in derivatives, it is given by (see Appendix \ref{constructionofactions} for details of the construction)
\be
S_\pi = M_\pi^2\int \rd^4 x\sqrt{-g}\left(-\frac{1}{2}(\partial\pi)^2+2H^2\pi^2-\pi(\partial\pi)^2+4H^2\pi^3\right)~,
\label{picubicaction}
\ee
which is consistent with~\eqref{goldstonequadraticaction} at quadratic order. This action non-linearly realizes time translations $P^0$, boosts $J^{0i}$ and the temporal component of special conformal transformations $K^0$, which act as
\bea
\nonumber
\delta_{P_0}\pi &=& \frac{1}{t } -\partial_t \pi \;; \\
\delta_{J_{0i}}\pi &=& \frac{x_i}{t } + t \partial_i \pi -x_i\partial_t \pi\;; \\
\nonumber
\delta_{K_0}\pi &=& -\frac{\vec x^2}{t } -\left(2t  x^\nu\partial_\nu-x^2\partial_t \right)\pi~.
\label{pitrans}
\eea
Notice that the --- na\"ively unstable --- growing mode solution $\pi \sim 1/t$ of \eqref{goldstonequadraticaction} corresponds merely to a non-linearly realized time translation, and is therefore harmless.

\section{SO(4,2) $\to$ SO(4,1) consistency relation}
\label{derivationoftheconsistencyrelation}

The next question is: {\it how do these non-linearly realized symmetries act on correlation functions?} In this Section we show that the non-linear realization of conformal symmetry constrains the form that correlation functions take in the limit that one of the $\pi$ external legs is taken to be very soft. 

\subsection{Derivation of the consistency relation}

In the case of inflation in the decoupling limit, the isometries of de Sitter are spontaneously broken by the inflaton's time-dependent background to the subgroup of rotations and translations. As a consequence of this spontaneous breaking, there are specific relations between correlation functions of different order. In particular, the $(N+1)$-point correlation functions in the squeezed limit are related to the variation of the $N$-point correlation functions under the broken symmetries (dilations and special conformal transformations). These relations go by the name of {\it consistency relations}~\cite{Maldacena:2002vr,Creminelli:2012ed,Creminelli:2004yq,Senatore:2012wy, Creminelli:2011rh, Cheung:2007sv}. They are the Ward identities resulting from the non-linearly realized symmetries in the broken phase of the theory~\cite{Assassi:2012zq,LamKurt,Goldberger:2013rsa}.

Our aim is to show that similar relations hold in the case of the nonlinearly-realized $SO(4,2)$ symmetries. We again expect that the squeezed limit of an $(N+1)$-point correlation function is related to the action of the broken generators on the $N$-point function. In this case the broken generators are time translations $P^0$, boosts $J^{0i}$, and the time component of a special conformal transformation $K^0$; correspondingly, the consistency relation will contain three pieces. 

In what follows we are going to use the background-wave arguments developed in~\cite{Creminelli:2012ed,Creminelli:2004yq,Senatore:2012wy}. In order to derive the consistency relation, the crucial insight is that, due to non-linear realization of $SO(4,2)$, the effect of a long background mode of $\pi$ can be obtained by a coordinate transformation
\be
\langle\phi(x_1)\ldots \phi(x_N) \rangle_{\pi_{\rm L}} = \langle\phi(\tilde{x}_1)\ldots \phi(\tilde{x}_N) \rangle \;,
\ee
where $\phi$ is a generic scalar field of the theory. As we will argue shortly, the change of coordinates is related to the value of the long mode as $\pi_{\rm L} = - \delta t  / t $. We can use  the previous equality to express the $(N+1)$-point correlation function in terms of the $N$-point function
\be
\langle \pi_{\rm L}(x)\phi(x_1)\ldots \phi(x_N) \rangle = \langle \pi_{\rm L}(x) \delta\langle\phi(x_1)\ldots \phi(x_N) \rangle_{\pi_{\rm L}} \rangle = \langle \pi_{\rm L}(x) \delta\langle\phi(\tilde{x}_1)\ldots \phi(\tilde{x}_N) \rangle \rangle \;.
\ee
This relation among correlation functions of different order in real space is the starting point of our derivation. By computing explicitly the variation of the $N$-point function on the right-hand side and then going to momentum space, we will obtain the consistency relation.

Before diving into the details of the full calculation, as an illustration of the method we consider the simplest case where we have only broken time translation. The coordinate transformation
\be
t \mapsto t '=t +a_0
\ee
will induce the change in the perturbation around the background $\pi'(x')=\pi(x) - a_0/t $, thus generating the homogeneous mode
\be
\pi_{\rm L} =- \frac{a_0}{t } \;.
\label{a0t}
\ee
On the one hand, the $(N+1)$-point function satisfies
\be
\langle \pi_{\rm L}\phi_1 \ldots \phi_N \rangle = \langle \pi_{\rm L} \delta\langle\phi_1 \ldots \phi_N \rangle_{\pi_{\rm L}} \rangle \;.
\label{N+1reln}
\ee
On the other hand, the $N$-point function in the presence of a long mode is the same as the $N$-point function without the long background but computed in the transformed coordinates:
\be
\delta\langle\phi_1\ldots \phi_N\rangle_{\pi_{\rm L}} = \delta\langle\phi_1 \ldots \phi_N\rangle _{t \rightarrow t +a_0} = a_0\frac{\rmd }{\rmd t }\langle\phi_1 \ldots \phi_N \rangle \;.
\ee
Using the relation~(\ref{a0t}) between $\pi_{\rm L}$ and $a_0$, and substituting the result in~(\ref{N+1reln}), we finally obtain
\be
\langle \pi_{\rm L}\phi_1 \ldots \phi_N \rangle = - \langle \pi_{\rm L}\pi_{\rm L} \rangle  t  \frac{\rmd}{\rmd t }\langle \phi_1 \ldots \phi_N \rangle \;.
\ee

This relation is written in real space but going to momentum space is trivial: a homogeneous long mode translates to the zero-momentum limit. Thus the $(N+1)$-point function has to be evaluated in the squeezed configuration
\be
\label{dilationscons}
\langle \pi_{\vec q}\;\phi_{\vec k_1}\ldots \phi_{\vec k_N} \rangle' _{\vec q \rightarrow 0}= - \langle \pi_{\vec q} \pi_{\vec q} \rangle'  t  \frac{\rmd}{\rmd t }\langle \phi_{\vec k_1} \ldots \phi_{\vec k_N} \rangle' \;,
\ee
where primes indicate that the factor of $(2\pi)^3\delta(\sum \vec k_i)$ has been removed from correlation functions. The previous expression has the same form as Maldacena's consistency relation for standard inflation \cite{Maldacena:2002vr,Creminelli:2004yq}. This is not a surprising result. In the case of the decoupling limit of standard inflation, Maldacena's consistency relation can be seen as a consequence of broken dilation invariance. The squeezed limit of the $(N+1)$-point function is related to the scale transformation of the $N$-point function. In our case, where time translations are spontaneously broken, we have a similar results --- the squeezed limit of the $(N+1)$-point function with a soft $\pi$ is related to the variation in time of the $N$-point function.

We are now ready to derive the most general consistency relation by applying the above method to the full
$SO(4,2) \rightarrow SO(4,1)$ symmetry breaking pattern. Since we have 5 broken symmetries but
only one Goldstone, the consistency relation will contain several terms of different order in $q$, each
coming from a different broken generator. Our starting point is the most general infinitesimal conformal transformation of space-time, which can be written as
\bea
\nonumber
t \mapsto t '&=&t +a_0 + c_ix_i + (\lambda -1) t +2(b_\mu x^\mu)t  - b_0(-t ^2+\vec x^2) \equiv t  +\delta t  \;, \\
x_i\mapsto x_i'&=& x_i+a_i - c_it +J_{ij}x_j + (\lambda -1)  x_i + 2(b_\mu x^\mu)x_i - b_i(-t ^2+\vec x^2) \equiv x_i +\delta x_i \;.
\eea
Notice that in $\delta t $ and $\delta x_i$, we can neglect all ${\cal O}(t)$ and ${\cal O}(t^2)$ terms because these will give subdominant contributions in the late-time limit. The remaining terms in $\delta x_i$ consist of  time-independent translation, rotation, dilation and $3d$ special conformal transformation, all of which are linearly realized symmetries. We conclude that for the purpose of deriving the consistency relation the only relevant part of the coordinate transformation is the time transformation:
\be
t \mapsto t '=t +a_0 + c_ix_i - b_0\vec x^2 = t  +\delta t  \;.
\label{tcoord}
\ee
This induces a long mode $\pi_{\rm L}$ of the form
\be
\pi_{\rm L}(x) = -\frac{\delta t }{t } = -\frac{1}{t }(a_0 + c_ix_i - b_0\vec x^2) \;.
\label{piLgen}
\ee
Two important points are worth emphasizing about the shape of the long mode. First, unlike the case when only the time-translation is broken, the long mode is in general $x$-dependent. Three different broken generators correspond to three terms with different powers of $x_i$. At the level of the consistency relation, this will translate to different broken generators constraining different powers of $q$ in the $\vec q \rightarrow 0$ expansion of correlation functions. Second, the above $\pi_{\rm L}$ does not correspond to the most general shape of a mode expanded to second order
\be
\pi(x) = \pi(0)  + \partial_i\pi(0)  x^i + \frac{1}{2} \partial_i\partial_j\pi(0)\ x^ix^j +\ldots\;
\ee
Specifically, our coordinate transformation induces only an {\it isotropic} profile for the long mode at quadratic order in coordinates (proportional to $\vec x^2$), while the traceless part of $\partial_i\partial_j\pi(0)$
cannot be generated through the coordinate transformation. Consequently, the consistency relation can only inform us about quadratic terms that are averaged over all angles. We focus only on the trace part of the long mode
\be
\pi_{\rm L}(x) \overset{\rm avg}{=} \pi_{\rm L}(0) + \partial_i\pi_{\rm L}(0)x^i + \frac{1}{6} \partial^2\pi_{\rm L}(0) \vec x^2 \,,
\ee
where here $\overset{\rm avg}{=}$ means that we have averaged over angles for the terms which are second order in gradients. Comparing this profile with~(\ref{piLgen}) we can read off the values of $a_0$, $c_i$ and $b_0$:
\be
\label{parameters}
a_0= -t \pi_{\rm L}(0) \;; \quad c_i = -t \partial_i \pi_{\rm L}(0) \;; \quad b_0 = \frac{1}{6}t\partial^2\pi_{\rm L}(0) \;.
\ee

We now have at our disposal all the necessary ingredients to derive the full consistency relation. The first step consists of computing the variation of an $N$-point function under the coordinate transformation~(\ref{tcoord}). Since the only relevant part of the full conformal transformation is a spatially-inhomogeneous shift in time, it is natural to first calculate unequal-time correlation functions and then take the equal-time limit at the end of the derivation. As another technical step, we must choose a point around which to perform our coordinate transformation. Since the fields in the correlation function are evaluated at $\vec x_1,\ldots, \vec x_N$, a natural choice is their average location:
\be
\vec x_+\equiv \frac{1}{N}\sum_{a=1}^N \vec x_a\,. 
\ee
In this way the origin of coordinate transformation is translated in the region of space where the correlators are calculated. The choice of $\vec x_+$ is of course arbitrary, and will drop out of the final result. The upshot is that the coordinate transformation of interest is
\be
t \rightarrow t ' = t +a_0^+ -c_i^+(x_i^+-x_i) - b_0^+(\vec x_+-\vec x)^2 \;.
\ee
where the coefficients are of the form \eqref{parameters}, but now evaluated at $x_+$: $a_0^+ = -t\pi_{\rm L}(x_+)$, $c_i^+ = -t\partial_i\pi_{\rm L}(x_+)$ and $b_0^+ = t\partial^2\pi_{\rm L}(x_+)/6$.

In order to be consistent, we must average all quadratic terms --- as a result, the $2b_0^+\vec x \cdot \vec x_+$ term will give zero contribution. The variation of an arbitrary $N$-point function under this coordinate transformation is
\be
\delta\langle\phi(t_1,\vec x_1)\ldots \phi(t _N,\vec x_N) \rangle_{\pi_{\rm L}} \overset{\rm avg}{=} \sum_{a=1}^N \big(a_0^+ -c^+_i(x_i^+ - x_{ai})- b^+_0( \vec x_+^2 + \vec x_a^2)\big)\frac{\partial }{\partial t_a}\langle\phi(t_1,\vec x_1)\ldots \phi(t_N,\vec x_N) \rangle \;.
\ee
Rewriting this equation in momentum space and using~\eqref{parameters}, we obtain
\bea
& &  \delta \langle \phi(t _1,\vec x_1)\ldots \phi(t _N,\vec x_N) \rangle_{\pi_{\rm L}} \overset{\rm avg}{=}  \int \frac{\rmd^3\vec k_1}{(2\pi)^3}\ldots \frac{\rmd^3\vec k_N}{(2\pi)^3} \frac{\rmd^3\vec q_+}{(2\pi)^3}(2\pi)^3 \delta^3(\vec P) \pi_{\vec q_+}(t )\\\nonumber
&&\times  \sum_a \left( -t  \frac{\partial}{\partial t_a}{\mathcal A(t_1,\ldots,t_N)} \right)  \(1 -iq_i^+(x_i^+ -x_{ai}) - \frac{1}{6}\vec q_+^{\;2} (\vec x_+^2  + \vec x_a^2) \) e^{i\vec x_+ \cdot \vec q_+}e^{i\sum \vec k_b \cdot \vec x_b},
\eea
where $\vec P \equiv \vec k_1 + \ldots + \vec k_N$ is the total momentum and $\mathcal{A}$ is the Fourier space $n$-point function. Expanding the phase $e^{i\vec x_+ \cdot \vec q_+}$
in small momentum $\vec q_+$ and averaging all second order terms, we get
\be
 \( 1 - iq_i^+x_i^+ - \frac{1}{6}\vec q_+^{\;2}\vec x_+^2 +iq_i^+x_{ai}  - \frac{1}{6}\vec q_+^{\;2}\vec x_a^2\)e^{i\vec x_+ \cdot \vec q_+}  \overset{\rm avg}{=} 1 +iq_i^+x_{ai}  - \frac{1}{6}\vec q_+^{\;2}\vec x_a^2 + \mathcal{O}(q_+^3) \;.
\ee
The result is independent of $\vec x_+$, as anticipated. 

The final step in the derivation is to average over the long mode $\pi_{\rm L}(t,\vec x)$. After performing the average, integrating over $\vec q_+$ and replacing the coordinates in front of the exponent with derivatives with respect to the momenta, we obtain the $(N+1)$-point correlation function
\begin{align}
\label{consderivation}
\langle \pi(t,\vec x) & \phi(t_1,\vec x_1)\ldots \phi(t_N,\vec x_N) \rangle \overset{\rm avg}{=}  \int \frac{\rmd^3\vec k_1}{(2\pi)^3}\ldots \frac{\rmd^3\vec k_N}{(2\pi)^3} \frac{\rmd^3\vec q}{(2\pi)^3}(2\pi)^3 \delta^3(\vec P) P_\pi(q) \nonumber \\
& \times  \sum_a \left( -t \frac{\partial}{\partial t_a}{\mathcal A(t_1,\ldots,t_N)} \right)   \(1 - q_i\partial_{k_{ai}}  + \frac{1}{6}\vec q^{\;2}\partial_{k_a}^2 \) e^{i\vec q \cdot \vec x + i\sum \vec k_b\cdot \vec x_b}   \;.
\end{align}
Next we integrate by parts to move the $\partial_{k_{ai}}$ derivatives from the exponent to the amplitude $\mathcal{A}$. Since the delta function $\delta^3(\vec P)$ depends on the momenta $\vec k_1,\ldots, \vec k_N$, this will introduce new terms involving derivatives of the delta function, but these are in fact precisely what is needed to build up the Taylor expansion of  $\delta^3(\vec P + \vec q)$. Indeed, averaging all second order terms, it is easy to prove the following relation\footnote{Notice a change of the sign in the second term in the brackets coming from integrating by parts.}
\be
\(1 + q_i\partial_{k_{ai}}  + \frac{1}{6}\vec q^{\;2}\partial_{k_a}^2 \) \left( \delta^3(\vec P) \mathcal A'  \right) \overset{\rm avg}{=} \delta^3(\vec P + \vec q) \(1 + q_i\partial_{k_{ai}}  + \frac{1}{6}\vec q^{\;2}\partial_{k_a}^2 \) \mathcal A' \;.
\ee
Going to momentum space on the left-hand side of~\eqref{consderivation} and comparing the integrands on each side of the equation, we obtain
\be
\label{resumed1}
\langle \pi_{\vec q}(t)\phi_{\vec k_1}(t_1)\ldots \phi_{\vec k_N}(t_N) \rangle'_{q\rightarrow 0} \overset{\rm avg}{=} -t P_\pi(q)\sum_a \(1 + q_i\partial_{k_{ai}}  + \frac{1}{6}\vec q^{\;2}\partial_{k_a}^2 \)  \frac{\partial}{\partial t_a} \langle \phi_{\vec k_1}(t_1)\ldots \phi_{\vec k_N} (t_N)\rangle' \;.
\ee
As expected, the $(N+1)$-point correlation function in the squeezed limit is related to the action of broken generators of the conformal algebra on the $N$-point function. Notice that on both sides the same delta function $\delta^3(\vec P + \vec q)$ has been removed from correlation functions, which implies in particular that the $N$ momenta on the right-hand side do not form a closed polygon. As we will see later with explicit checks, this prescription is essential for the consistency relation to hold.

From~(\ref{resumed1}), it is easy to deduce the equal-time version of the consistency relation:
\be
\label{resumed2}
\langle \pi_{\vec q}\phi_{\vec k_1}\ldots \phi_{\vec k_N} \rangle'_{q\rightarrow 0} \overset{\rm avg}{=} -P_\pi(q)\(1 + \frac{1}{N} q_i \sum_a \partial_{k_{ai}}  + \frac{1}{6N}\vec q^{\;2} \sum_a \partial_{k_a}^2 \)  t \frac{\rmd}{\rmd t} \langle \phi_{\vec k_1}\ldots \phi_{\vec k_N} \rangle' \;.
\ee
This is the main result of the paper. In the remainder of this Section, we will verify the consistency relation in a few examples. Explicit calculations of the correlation functions will be given in Appendix~\ref{correlationfncomputation}. 

\subsection{Explicit checks of the consistency relation}

The simplest checks involve the squeezed limit of a three-point function. As a general rule, note that the three-point function in the squeezed limit does not have any linear correction in $\vec q$. Indeed, using the symmetric expansion of $\vec k_1$ and $\vec k_2$,
\be
\label{3pfmomenta}
\vec k_1 = \vec k- \frac{\vec q}{2}\;; \quad \vec k_2 = -\vec k - \frac{\vec q}{2}\;; \quad \vec k_1+\vec k_2 +\vec q = 0\;,
\ee
it is easy to see that~\eqref{resumed2} does not give a linear term in $q$. Consider the three-point function of a $\pi$ with two massless spectator field $\chi$ given by~\eqref{3pfmassless}. Its expansion in the soft limit,
\be
\label{pichichisq}
\langle\pi_{\vec q}(t) \chi_{\vec k_1}(t) \chi_{\vec k_2}(t) \rangle'_{q\rightarrow 0} = \frac{3\pi H^4}{16M_\pi^2M_\chi^2}\frac{1}{q^5k^6 t } \left( 3 (\vec k\cdot \vec q)^2  - k^2q^2 + \mathcal{O}(q^3) \right) \;,
\ee
starts at order $q^2$. Notice that the angular average of the ${\cal O}(q^2)$ terms vanishes, in agreement with the consistency relation~(\ref{resumed2}) ---
since the two-point function $\langle \chi_{\vec k_1}\chi_{\vec k_2}\rangle'$ is independent of time for massless fields, the right-hand side of~\eqref{resumed2} vanishes in this case. 


As a less trivial check, consider the three-point function of $\pi$ with massive ($\Delta=1$) fields $\vp$ given by~(\ref{piphiphiappendix}). Its squeezed limit after averaging is
\be
\langle\pi_{\vec q}\vp_{\vec k_1}\vp_{\vec k_2}\rangle'_{q\rightarrow 0} = -\frac{9H^4}{2M_\vp^2M_\pi^2}\frac{1}{q^5k}\left( 1+ \frac{3 (\vec k\cdot \vec q)^2  - k^2q^2}{4k^4} \right) \overset{\rm avg}{=} -\frac{9H^4}{2M_\vp^2M_\pi^2}\frac{1}{q^5k} \;.
\ee
It is straightforward to verify that the consistency relation is satisfied at zeroth order in $q$ using the explicit expressions for the power spectra given by~\eqref{pipi} and~\eqref{phiphi}. Similarly to the massless case, the ${\cal O}(q^2)$ terms average to zero. On the other hand, from the right-hand side of~(\ref{resumed2}) we conclude that the consistency relation seemingly predicts non-vanishing ${\cal O}(q^2)$ corrections. This is an illusion: remembering that the hard momenta do not form a closed polygon, $\vec k_1$ and $\vec k_2$ are not equal in magnitude and opposite, and using~(\ref{3pfmomenta}) we obtain
\be
\frac{1}{k_{1;2}} = \frac{1}{k} \left( 1 \pm \frac{\vec k \cdot \vec q}{2k_{1;2}^2} + \frac{3(\vec k \cdot \vec q)^2 - q^2k^2}{8k^4} \right) \;.
\ee
Averaging, this gives
\bea
\nonumber
\frac{1}{\sqrt{k_1k_2}} &\overset{\rm avg}{=}& \frac{1}{k} \left( 1 - \frac{1}{24}\frac{q^2}{k^2} \right) \;;\qquad \frac{1}{2}q_i\sum\partial_{k_{ai}} \frac{1}{\sqrt{k_1k_2}} \overset{\rm avg}{=}  \frac{1}{k} \left( \frac{1}{12}\frac{q^2}{k^2} \right) \;; \\
\frac{1}{12} q^2 \sum_a \partial_{k_a}^2 \frac{1}{\sqrt{k_1k_2}} &\overset{\rm avg}{=}& \frac{1}{k} \left( -\frac{1}{24}\frac{q^2}{k^2} \right) \;.
\eea
Thus, we see that all ${\cal O}(q^2)$ contributions to the right-hand side of~(\ref{resumed2}) cancel, as they should. 

Similar relations can be used to check the consistency relation for the three-point function of the Goldstone field $\pi$, given by~(\ref{pi3ptfn}). Its squeezed limit (after averaging) gives
\be
\langle\pi_{\vec q}\pi_{\vec k_1}\pi_{\vec k_2}\rangle'_{q\rightarrow 0} = \frac{81H^4}{2M_\pi^4 t^4}\frac{1}{q^5k^5}\left( 1+ \frac{5}{8}\frac{7 (\vec k\cdot \vec q)^2 - k^2q^2}{k^4} \right) \overset{\rm avg}{=} \frac{81H^4}{2M_\pi^4 t^4}\frac{1}{q^5k^5} \left( 1 + \frac{5}{6}\frac{q^2}{k^2} \right) \;.
\ee
Meanwhile, expanding the right-hand side of~(\ref{resumed2}) in the soft momentum $\vec q$ gives
\be
\frac{81H^4}{2M_\pi^4 t^4}\frac{1}{q^5} \left( 1 + \frac{1}{2}q_i\sum\partial_{k_{ai}} + \frac{1}{12} q^2 \sum_a \partial_{k_a}^2 \right)\frac{1}{\sqrt{k_1^5k_2^5}} \overset{\rm avg}{=}\frac{81H^4}{2M_\pi^4 t^4}\frac{1}{q^5k^5} 
 \left( 1 + \frac{5}{6}\frac{q^2}{k^2} \right) \;.
\ee
The consistency relation is satisfied.

We can also check the consistency relation for the squeezed limit of the four-point function~\eqref{4pfpichichichi} with one Goldstone and three massive ($\Delta=1$) fields $\vp$. In this case we can express one of the momenta, say $\vec k_3$, in terms of the others: $\vec k_3 = - (\vec k_1+\vec k_2) - \vec q$. Averaging all terms proportional to $q^2$, we obtain the following set of relations
\bea
\nonumber
\frac{1}{k_3} &\overset{\rm avg}{=}&\frac{1}{p}\left( 1 + \frac{\vec p\cdot \vec q}{p^2} \right)\;;\quad \frac{1}{|\vec k_i +\vec q|} \overset{\rm avg}{=} \frac{1}{k_i}\left( 1 - \frac{\vec k_i \cdot \vec q}{k_i^2} \right)~~(i=1,2)\;;\\
 \frac{1}{|\vec k_3 +\vec q|} &=& \frac{1}{p}\;; \qquad \;\;\;\;\;\;\;\;\;\;\;\;\;\;\;\;\;\;\;\;\;\;\;\;\; k_3 \overset{\rm avg}{=} p \left( 1 + \frac{\vec p\cdot \vec q}{p^2} + \frac{1}{3} \frac{q^2}{p^2}\right)\;,
\label{k3expand}
\eea
where $\vec p \equiv \vec k_1 + \vec k_2$. Using these equations, the four-point function can be expanded to ${\cal O}(q^2)$ as
\be
\label{squeezed4pf}
\langle\pi_{\vec q}\vphi_{\vec k_1}\vphi_{\vec k_2} \vphi_{\vec k_3} \rangle'_{q\rightarrow 0} \overset{\rm avg}{=} -\frac{81\pi H^4\lambda t}{8M_\pi^2M_\vp^4} \frac{1}{q^5k_1k_2p}\left(1- \frac{\vec k_1\cdot \vec q}{3k_1^2} - \frac{\vec k_2\cdot \vec q}{3k_2^2} - \frac{2\vec p \cdot \vec q}{3p^2} + \frac{2}{9}\frac{q^2}{p^2} \right)\;.
\ee
Meanwhile, from the right-hand side of the consistency relation we are instructed to act with various derivatives on the three-point function for massive fields (see~\eqref{vpvpvp}):
\be
\langle\vp_{\vec k_1}\vp_{\vec k_2}\vp_{\vec k_3}\rangle' = \frac{3\pi H^2\lambda}{4M_\vp^4}\frac{t^3}{k_1k_2k_3}\;.
\ee
Noting that the action of the Laplacian on monopole terms gives zero, $\partial^2_{k_a}\frac{1}{k_a}=0$, and taking into account the power spectrum of the long mode $\pi$, the right-hand side of the consistency relation becomes
\bea
\nonumber
\mbox{r.h.s.~of~(4.23)} &=& -\frac{81\pi H^4\lambda t}{8M_\pi^2M_\vp^4} \frac{1}{q^5k_1k_2 k_3}\left(1- \frac{\vec k_1\cdot \vec q}{3k_1^2} - \frac{\vec k_2\cdot \vec q}{3k_2^2} - \frac{\vec k_3\cdot \vec q}{3k_3^2} \right)\\
&\overset{\rm avg}{=}& -\frac{81\pi H^4\lambda t}{8M_\pi^2M_\vp^4} \frac{1}{q^5k_1k_2p}\left(1- \frac{\vec k_1\cdot \vec q}{3k_1^2} - \frac{\vec k_2\cdot \vec q}{3k_2^2} - \frac{2\vec p \cdot \vec q}{3p^2} + \frac{2}{9}\frac{q^2}{p^2} \right)\;.
\eea
where in the last step we have used the expansion of $\vec k_3$. This verifies the consistency relation in this case.

\section{Connection to observables: soft internal lines and anisotropy of the power spectrum}
\label{softlinesrubakov}

As we discussed, the breaking of $SO(4,2)$ implies the existence of the Goldstone field $\pi$, and consequently the consistency relation we derived constrains correlation functions with soft external Goldstone lines. Unfortunately, the cosmological perturbations we observe come from a spectator field and not from $\pi$, so that it is not obvious how one can connect the previous results to observations. There are, however, two situations in which $SO(4,2)$ is observationally relevant. The first is when diagrams of the spectator field contain soft {\em internal} $\pi$ lines (for similar results in inflation see \cite{Seery:2008ax,Leblond:2010yq,Creminelli:2012ed}). Internal soft $\pi$ lines are expected to give the dominant contribution when a sum of external momenta becomes small, and they will dominate in comparison with soft internal lines of the spectator field, because of the very red spectrum of the Goldstone. The second possibility stems from the fact that, even if $\pi$ is not directly measured, its value during the conformal phase is correlated with the modes of the spectator field and thus changes their statistics. In particular, very long modes of $\pi$ induce an anisotropy in the spectator field power spectrum. These two observational features were studied in \cite{Libanov:2010nk,Libanov:2010ci, Libanov:2011bk}. Here we want to stress that these properties are a direct consequence of the non-linear realization of $SO(4,2)$ and not specific to a given model. We will also find an additional important contribution to the four-point function from a loop of $\pi$ fields that has been overlooked in the literature. This contribution may be larger than the tree-level $\pi$ exchange and it is phenomenologically quite different. 

\begin{figure}
\centering
\includegraphics[width=3.7in]{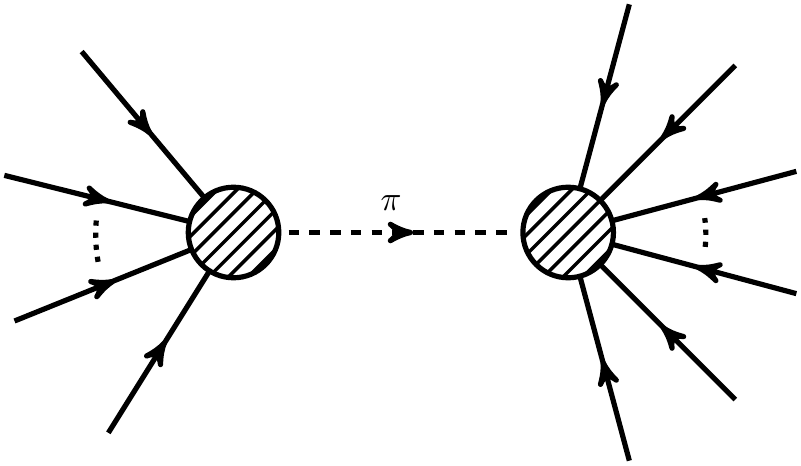}
\caption{\small\label{fig:softinternal} Factorization of an $(N+M)$-point function via an exchange of a soft $\pi$.}
\end{figure}

Let us start with soft internal $\pi$ lines. In the limit in which the sum of $N$ external momenta becomes small, the amplitude of an $(N+M)$-point function factorizes in the following way (see Fig.~\ref{fig:softinternal})
\be
\label{factor}
\langle \chi_{\vec k_1}\ldots \chi_{\vec k_{M+N}} \rangle'_{q\rightarrow 0} =\frac{1}{P_\pi(q)} \langle \pi_{-\vec{q}} \chi_{\vec k_1}\ldots \chi_{\vec k_M} \rangle'_{q\rightarrow 0} \langle \pi_{\vec{q}} \chi_{\vec k_{M+1}}\ldots \chi_{\vec k_{M+N}} \rangle'_{q\rightarrow 0} \,.
\ee
The $(N+1)$ and $(M+1)$-point functions are severely constrained by the $SO(4,1)$ symmetry and their squeezed limit is further constrained by the non-linear realization of $SO(4,2)$. In this way, the amplitude for the $(N+M)$-point function with a soft internal line can be expressed in terms $N$ and $M$-point functions. 
The simplest case is the four-point function of massless spectator fields, which was studied in detail in \cite{Libanov:2010nk, Libanov:2010ci, Libanov:2011bk}. Using the factorization \eqref{factor} above and the squeezed limit~\eqref{pichichisq} for the three-point function $\langle\pi \chi\chi\rangle$, in the limit $\vec k_1 + \vec k_2 \equiv \vec q \to 0$ we get
\be
\label{Ruby}
\langle \chi_{\vec k_1}\ldots \chi_{\vec k_4} \rangle'_{q\rightarrow 0} =\frac{\pi^2}{144} {\cal P}_\pi {\cal P}_\chi^2 \frac1{q k_1^4 k_3^4} \left( 3 (\hat k_1\cdot \hat q)^2  - 1  \right) \left( 3 (\hat k_3\cdot \hat q)^2  - 1 \right)   \;,
\ee
where ${\cal P}_\pi \equiv 9H^2/2M_\pi^2$~; ${\cal P}_\chi \equiv H^2/2M_\chi^2$ are the dimensionless power spectra.
It is important to stress that the shape of the four-point function in the soft internal limit is completely specified by symmetries since the three-point function $\langle\pi \chi\chi\rangle$ is completely fixed by $SO(4,1)$ up to an overall constant. Notice that the squeezed limit of the three-point function is constrained, as we discussed in the previous Section, by $SO(4,2)$ as well. In the massless case we cannot obtain terms scaling as $q^0$ or $q^1$, and all terms scaling as $q^2$ must vanish when averaged over the angles. This is indeed what we have in \eqref{Ruby}.

The four-point function becomes very large in the $q \to 0$  limit, as it scales as $1/q$. This is a consequence of the very red spectrum of $\pi$ and it can be contrasted, for example, with inflationary models with reduced speed of sound which are regular in the $q \to 0$ limit \cite{Chen:2009bc,Arroja:2009pd} . We conclude that {\em a four-point function which becomes large in the soft internal (collapsed) limit, with the precise shape \eqref{Ruby}, is a general prediction of the conformal mechanism}. Notice, however, that the overall multiplicative constant in~\eqref{Ruby} cannot be fixed by symmetry arguments.  

If one assumes a linear relation between $\zeta$ and $\chi$ (non-linearities will give additional model-dependent contributions to local non-Gaussianity) we get that the four-point function above has an amplitude
\be
\frac{\langle\zeta\zeta\zeta\zeta\rangle}{{\cal P}_\zeta^3} \simeq
\frac{\pi^2}{144} \cdot \frac{{\cal P}_\pi}{{\cal P}_\zeta}  \;.
\ee
Although data analysis has not been performed for the particular momentum dependence of \eqref{Ruby}, one can get a rough constraint using limits on equilateral models of four-point function\footnote{This may be slightly conservative as equilateral shapes are regular in the limit $q \to 0$.} obtained in \cite{Fergusson:2010gn}: $|t_{\rm NL}^{\rm equil}| \lesssim 7 \cdot 10^6$. This gives
\be
\label{eq:4pftree}
{\cal P}_\pi \lesssim 500 \;.
\ee

The four-point function we studied is obtained by averaging over the long wavelength modes of $\pi$. However, if we do not take the statistical average, we still have a realization-dependent effect: long modes of $\pi$ induce an anisotropy in the power spectrum of the short modes, as pointed out in~\cite{Libanov:2010nk, Libanov:2010ci, Libanov:2011bk}. Notice that this is possible even though $\pi$ does not contribute to the observed perturbations: its value during the conformal phase still affects the observable modes of the spectator field. This effect is also completely fixed, up to an overall constant, by the symmetries of the problem. 

The effect of a long $\pi$ mode on the observable 2-point function can be read from the three-point function $\langle \pi\chi\chi\rangle$, given by~\eqref{pichichisq}, in the squeezed limit
\be
\langle\pi_{\vec q} \chi_{\vec k_1} \chi_{\vec k_2} \rangle'_{q\rightarrow 0} = \frac{\pi}{12} {\cal P}_{\pi} {\cal P}_{\chi} \frac{1}{q^5k^6 t} q^2k^2 \left( 3 \cos^2 \theta  - 1 \right) \;,
\ee 
where $\theta$ is the relative angle between $\vec k$ and $\vec q$.
 We can write the variation of the power spectrum of $\chi$ in the presence of a given realization of the $\pi$ field in a schematic way as
\be
\delta \langle \chi_{\vec k} \chi_{-\vec k} \rangle' = \frac{\langle \pi_{\vec q} \chi\chi\rangle'_{q\to 0}}{\langle \pi_{\vec q}\pi_{-\vec q} \rangle'_{q\to 0}} \pi_{\vec q}  =  \langle \chi_{\vec k} \chi_{-\vec k} \rangle' \cdot \frac{\pi}{12} \frac{1}{k} (3\cos^2\theta -1) t q^2\pi_{\vec q} \;.
\ee
All modes $\pi_{\vec q}$ which are outside the present Hubble radius will contribute to the anisotropy of the $\chi$ power spectrum. The typical size of the effect is given by the square root of the variance calculated by summing over all super Hubble modes
\be
\int \frac{{\rm d}^3q}{(2 \pi)^3} \; \langle t^2 q^4 \pi_{\vec q}\pi_{-\vec q} \rangle' = \frac1{2\pi^2}\int_0^{H_0} q^2 \rmd q\; t^2 q^4 \frac{{\cal P}_\pi}{q^5t^2} \sim  \frac1{4\pi^2} {\cal P}_{\pi} H_0^2\;.
\ee
This gives
\be
\label{anisotropy} 
\langle \chi_{\vec k} \chi_{-\vec k} \rangle'_{\pi}=\langle \chi_{\vec k} \chi_{-\vec k} \rangle'\left( 1 + c_1 \frac{\sqrt{{\cal P}_\pi}}{2 \pi} \frac{H_0}{k}(3 \cos^2 \theta -1 )  \right) \;,
\ee
where $c_1$ is a number of order unity, which depends on our position in the Universe \cite{Libanov:2010nk, Libanov:2010ci, Libanov:2011bk}.

Another source of anisotropy in the power spectrum arises by considering a four-point function $\langle \pi\pi\chi\chi \rangle$  \cite{Libanov:2010nk, Libanov:2010ci, Libanov:2011bk}.  This induces a variation of the 2-point function $\langle \chi\chi \rangle$ in the presence of two long modes of $\pi$. In this case the $SO(4,2)$ symmetry fixes both the shape and the normalization of the effect.  The variation of the 2-point function $\langle \chi\chi \rangle$ in the presence of two long background modes $\pi_1$ and $\pi_2$ corresponds to the composition of the associated $SO(4,2)$ transformations. A possible issue is that the broken generators $K_0$, $J_{0i}$ and $P_0$ do not commute, so that the overall transformation seems to depend on the ordering.
Fortunately, in our case all the commutators of the broken generators of $SO(4,2)$ give {\em unbroken} generators. These do not change the 2-point function, so that we do not have to worry about non-commutativity in the case at hand. Since the 2-point function is time-independent, its variation at lowest order in gradients will come from a boost at second order. Without loss of generality, we consider a boost along the $x$-direction. The transformation of coordinates is given by
\be
\label{2boost}
x'=\gamma(x-v_x t)\,,\; y'=y\,,\; z'=z\,,\; t'=\gamma(t-v_x x)\;,
\ee
where $\gamma\equiv(1-v^2)^{-1/2}$. Neglecting parts proportional to $t$, the induced background field is, up to second order in $v_x$,
\be
\pi = -\frac{\delta t}{t} = \frac{v_x x}{t}\;.
\ee
In momentum space, the parameter $v_x$ is given by
\be
v_x = itq_x \pi_{\vec q}\;.
\ee
The transformation \eqref{2boost} implies that in momentum space $k_x$ component of the wavevector has to be multiplied by $\gamma^{-1}$, while $k_y$ and $k_z$ remain the same. Expanding $k^{-3}$ in the denominator of the power spectrum, we find that the effect on the 2-point function of $\chi$ is:
\be
\label{boostonchi}
\delta \langle \chi_{\vec k} \chi_{-\vec k} \rangle' = \langle \chi_{\vec k} \chi_{-\vec k} \rangle \frac{3}{2}\frac{(\vec v \cdot \vec k)^2}{k^2}=  -\langle \chi_{\vec k} \chi_{-\vec k} \rangle \frac{3}{2}t^2q^2\pi_{\vec q}^2  \cos^2 \theta \;.
\ee
We can calculate the typical value of $t^2q^2\pi_{\vec q}^2$ in a way similar to before:
\be
\int \frac{\rd^3q}{(2 \pi)^3} \;  \langle t^2 q^2 \pi_{\vec q}^2 \rangle = \frac{1}{2\pi^2} \int_0^{H_0} q^2 \rmd q\; t^2 q^2 \frac{{\cal P}_\pi}{q^5t^2} \sim \frac{1}{2\pi^2} {\cal P}_{\pi} \log\frac{H_0}{\Lambda}\;,
\ee
where $\Lambda$ is an IR cutoff. The contribution to the anisotropy is given by:
\be
\delta \langle \chi_{\vec k} \chi_{-\vec k} \rangle' =  -\langle \chi_{\vec k} \chi_{-\vec k} \rangle \frac3{4 \pi^2} {\cal P}_{\pi} \log\frac{H_0}{\Lambda}  \cos^2 \theta \;.
\ee
Combining with~(\ref{anisotropy}), the total anisotropy of the power spectrum is given by
\be
\label{totalanisotropy} 
\langle \chi_{\vec k} \chi_{-\vec k} \rangle'_{\pi_{\vec q}}=\langle \chi_{\vec k} \chi_{-\vec k} \rangle'\left( 1 + c_1 \frac{\sqrt{{\cal P}_\pi}}{2\pi} \frac{H_0}{k}(3 \cos^2 \theta -1 ) + c_2 \frac{3 \,{\cal P}_\pi }{4 \pi^2} \cos^2\theta \log \frac{H_0}{\Lambda} \right) \;,
\ee
where $c_2$ is another constant of order unity, which depends on the particular position in the Universe. The two sources of anisotropies are quite different. The first scales as $1/k$, and thus important only for long modes,
while the second is scale invariant. Moreover, the first contribution averages to zero if summed over the possible orientations between long and short modes, while the second does not. Notice also that the first effect is dominated by $\pi$ modes which are slightly longer than the present Hubble radius, while the second gets contributions from all scales as shown by the logarithmic dependence. The logarithmic enhancement can overcome the suppression due to the fact that the second effect is of order $\pi^2$ and not $\pi$.

\begin{figure}
\centering
\includegraphics[width=5.6in]{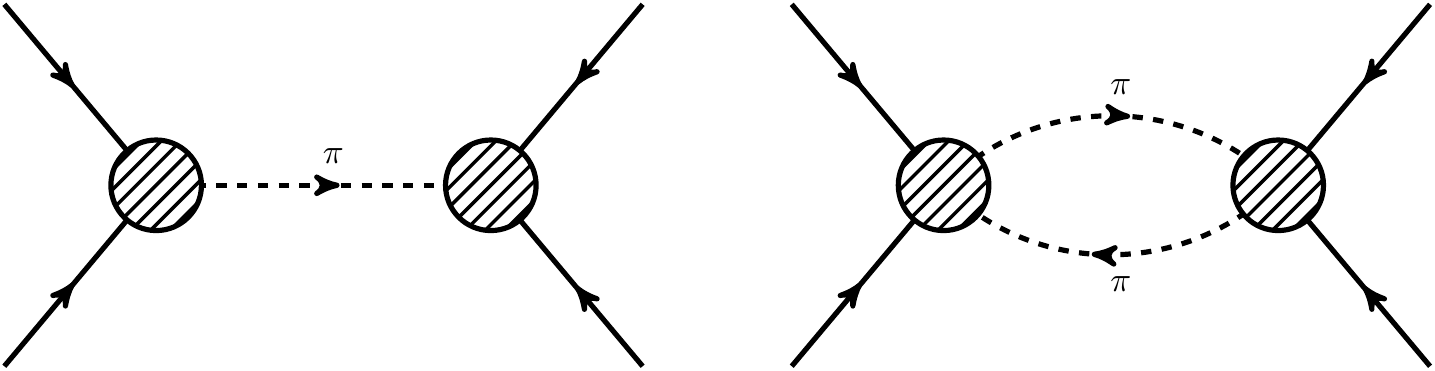}
\caption{\small \label{fig:onetwopi} The four-point function of $\chi$'s with an exchange of one and two soft $\pi$'s.}
\end{figure}

As we have seen, the second contribution to the power-spectrum anisotropy is related to the correlator $\langle\pi\pi\chi\chi\rangle$. This suggests that we missed a potentially large contribution to the four-point function of $\chi$'s in the soft internal limit, coming from a loop of soft $\pi$ particles (see Fig.~\ref{fig:onetwopi}). At first this looks worrisome as we expect a loop diagram to be small compared to a tree-level one. However, the situation is similar to the one we discussed for the anisotropy. When only one soft $\pi$ is exchanged, the interaction with the $\chi$'s arises at order $q^2$ as we discussed above. When two soft $\pi$'s are exchanged, on the other hand, each of them carries a single soft momentum, as the interaction arises from the non-linear realization of boosts. Therefore, in going from tree-level $\pi$ exchange to a one-loop diagram the number of $q$'s at the vertices remains the same, and we have the extra loop factor
\be
\int \frac{{\rm d}^3q}{(2\pi)^3} \frac{{\cal P}_\pi}{q^5} \sim \frac{{\cal P}_\pi}{q^2} \;.
\ee
If $q$ is small enough compared with the external momenta, the loop diagram will dominate over the tree level exchange. Notice that this does not signal a breakdown of perturbation theory: it is straightforward to check that the exchange of extra $\pi$'s is not further enhanced by powers of $1/q$, but only suppressed by powers of ${\cal P}_\pi$.

The loop diagram is straightforward to evaluate starting from \eqref{boostonchi}
\be
\label{oneloop}
\langle\chi_{\vec k_1}\chi_{\vec k_2}\chi_{\vec k_3}\chi_{\vec k_4}\rangle'_{q \to 0} = \frac92 \frac{{\cal P}_\chi}{k_1^3}\frac{{\cal P}_\chi}{k_3^3}  \int \frac{{\rm d}^3q_1}{(2\pi)^3} (\hat q_1 \cdot \hat k_1) (\hat q_2 \cdot \hat k_1) (\hat q_1 \cdot \hat k_3) (\hat q_2 \cdot \hat k_3)  \frac{{\cal P}_\pi}{q_1^3}  \frac{{\cal P}_\pi}{q_2^3} \;,
\ee
where $\vec q \equiv \vec k_1 + \vec k_2$ and $\vec q_1 + \vec q_2 = \vec q$. In writing this expression we have assumed that both internal legs are soft so that their coupling is fixed by the non-linear realization of $SO(4,2)$. Indeed we will see that the loop integral is dominated by having $q_1$ and $q_2$ both of order $q$. If we disregard the angular dependence and average over the direction of the short modes, we get
\be
\frac12 \frac{{\cal P}_\chi}{k_1^3}\frac{{\cal P}_\chi}{k_3^3}  \int \frac{{\rm d}^3q_1}{(2\pi)^3} (\hat q_1 \cdot \hat q_2)^2  \frac{{\cal P}_\pi}{q_1^3}  \frac{{\cal P}_\pi}{q_2^3} \;.
\ee
The loop integral is dominated by long modes and it is IR divergent, similarly to what happened for the anisotropy of the power spectrum. We get
\be
\label{eq:4pfloop}
\langle\chi_{\vec k_1}\chi_{\vec k_2}\chi_{\vec k_3}\chi_{\vec k_4}\rangle'_{q \to 0}\sim \frac{1}{24 \pi^2}\frac{{\cal P}_\chi}{k_1^3}\frac{{\cal P}_\chi}{k_3^3}   \frac{{\cal P}_\pi^2}{q^3} \log\frac{q}{\Lambda} \;.
\ee
As promised this result contains, when compared with the tree-level calculation~\eqref{Ruby}, a factor of ${\cal P}_\pi k^2/q^2$ which may be large for sufficiently small $q$. 

Notice that the momentum dependence of this result (after performing the angular average) is exactly the one of a $\tau_{\rm NL}$ non-Gaussianity. Again assuming a linear relation between $\zeta$ and $\chi$  we get
\be
\tau_{\rm NL} \sim \frac{1}{96 \pi^2}\frac{{\cal P}_\pi^2}{{\cal P}_\zeta} \log\frac{q}{\Lambda} \;.
\ee
Using the experimental limit $|\tau_{\rm NL}| \lesssim 2 \cdot 10^4$ \cite{Smidt:2010ra} and neglecting the logarithmic enhancement, one gets a rough limit on ${\cal P}_\pi$
\be
{\cal P}_\pi \lesssim {{\cal P}_\zeta}^{1/2} \cdot (96 \pi^2 \cdot 2 \cdot 10^4)^{1/2} \simeq 1\;.
\ee
This (rough) limit is stronger than the one obtained from the tree-level four-point function.
The four-point function \eqref{eq:4pfloop} will also contribute both to a stochastic scale-dependent bias \cite{Baumann:2012bc} and to the power spectrum of $\mu$-distortion \cite{Pajer:2012vz}. It would be interesting to understand whether the angle dependence, which is different from a standard $\tau_{\rm NL}$ shape, affects these observables. 

In this paper we only studied correlation functions in the absence of gravity. As discussed in \cite{Creminelli:2010ba,Libanov:2011zy}, this is a good approximation for sufficiently early times; $\pi$ perturbations give a negligible contribution to the observable quantity $\zeta$, while $\chi$ perturbations will source $\zeta$ by one of the standard conversion mechanisms.

An important concluding remark is in order. Our $SO(4,2)$ consistency relations are not as constraining as the ones for single-field inflation. In that case one can derive consistency relations directly in terms of the observed variable $\zeta$ which, if violated, would rule out any single-field model. Here, on other hand, we can just single out the effects due to the emission of $\pi$, but their relation with observables is ultimately model-dependent: for instance, all the effects we discussed vanish in the limit $P_\pi \to 0$. This is ultimately due to the fact that we are discussing a multi-field model, where perturbations are sourced by an isocurvature field. Even though we cannot derive completely model-independent relations, the red spectrum of $\pi$ makes the contributions discussed above sufficiently peculiar to be distinguishable from the other model-dependent effects.


\section{Subtleties}
\label{extrastuff}

We now address two important subtleties which arise in our analysis. The first (Sec.~\ref{assigningweights}) is that, due to the negative mass of the Goldstone field $\pi$, it is {\it prima facie} unclear when it should be possible to consistently assign $3d$ conformal weights for the transformations~\eqref{conformaltransdeltas}. We will see that this is only possible when the relevant interaction turns off sufficiently quickly, so that the in-in integral is dominated by early-time contributions.

The second subtlety (Sec.~\ref{offshell}) arises when checking the consistency relation: we must expand the correlation function slightly off-shell\footnote{Here, our usage of the phrase ``off-shell" is slightly nonstandard. What we mean is that the momenta in the correlation functions do not form a closed polygon.} to check the relations, but it is not clear that there is a unique way to do this. We show that not only is the procedure unambiguous, but it is possible to work on-shell at all times by rewriting the consistency relation at each order in $\vec q$ as a differential identity. These identities correspond precisely to the Ward identities derived from the broken symmetry generators~\cite{ward2}.

\subsection{Assigning conformal weights}
\label{assigningweights}
We first discuss a puzzle that arises when considering correlation functions involving the Goldstone field $\pi$. It should have conformal weight $\Delta=-1$. Indeed, the correlation function $\langle \pi\chi\chi \rangle$ with $\chi$ massless satisfies~\eqref{conformaltransdeltas} with the proper $\Delta$'s. On the other hand this is not the case for $\langle \pi\pi\pi\rangle$ or $\langle \pi\vp\vp\rangle$ with $\Delta_\vp=1$. These correlators {\em are} $SO(4,1)$ invariant but one is not allowed to replace the time dependence with the free-field conformal weights. 

The resolution of this apparent paradox is as follows: the possibility of assigning conformal weights to each field as a function of its mass depends on whether the late-time behavior of correlation functions is fixed by the free-field evolution. In other words, interactions must turn off sufficiently fast in time. Since one expects in general that the late-time evolution is purely classical, it is straightforward to ascertain by inspection whether conformal weights can be assigned. 

To make this a bit more concrete, consider the three-point function
\be
\langle \phi_1 (t)\phi_2 (t) \phi_3 (t)\rangle = - i \int^{t}_{- \infty (1 - i \epsilon)} \rd t' \langle \phi_1(t) \phi_2(t) \phi_3(t)  H_{\rm int}^{(3)}(t')\rangle + {\rm c.c.} 
\ee
where $H_{\rm int}^{(3)}$ is the cubic, interaction-picture Hamiltonian. The dependence on $t$ comes both from the integration limit and from the $\phi$ fields. This second contribution follows the free-field evolution and allows a consistent assignment of conformal weights. Thus, we want to study under which conditions the additional $t$ dependence from the integral can be neglected, {\it i.e.}, whether the integral is dominated by early times. 
Notice that the $i \epsilon$ prescription is different in the two complex-conjugated terms, but we can neglect this difference as we are only interested in the late-time behavior of the integral. Hence we can write
\be
\begin{split}
& \langle \phi_1 (t)\phi_2 (t) \phi_3 (t)\rangle  \sim - i \int^{t} \rd t' \;  \langle [ \phi_1(t) \phi_2(t) \phi_3(t) ,   H_{\rm int}^{(3)}(t') ] \rangle \\   = &- i \int^{t} \rd t' \;  \langle \phi_1(t) \phi_2(t) [  \phi_3(t) ,  H_{\rm int}^{(3)}(t') ] \rangle - i \int^{t} \rd t' \;  \langle \phi_1(t)  [  \phi_2(t) ,  H_{\rm int}^{(3)}(t') ]  \phi_3(t)\rangle \\ & - i \int^{t} \rd t' \;  \langle  [  \phi_1(t) ,  H_{\rm int}^{(3)}(t') ]  \phi_2(t)  \phi_3(t) \rangle \;.
\end{split} 
\ee
In each of the three terms, one leg is evolved non-linearly, and we want to show that this describes the classical non-linear evolution. Let us focus on the first term for concreteness. The commutator with the Hamiltonian will contain three pieces, depending on which of the three legs $\phi_3$ is paired with. For each of these contributions, we can integrate by parts to get rid of the derivatives acting on the leg we singled out and write the Hamiltonian as 
\be
H_{\rm int}^{(3)} = \int \rd^3 x  \; a^4 \; \phi(t') \; {\cal O}^{(2)}(t') \;,
\ee
where ${\cal O}^{(2)}(t')$ is a quadratic operator which will be paired with the legs $\phi_1$ and $\phi_2$. In the late-time limit, we can replace $\phi_1$ and $\phi_2$ within ${\cal O}^{(2)}(t')$ by their growing mode solutions,
which we take to be real. Thus ${\cal O}^{(2)}(t')$ becomes a number which acts like a source for the mode $\phi_3$. We have the mode expansion
\be
\phi(t') = \phi_{\rm cl} (t') \hat a^\dagger + \phi_{\rm cl}^* (t') \hat a~,
\ee
with $\phi_{\rm cl}$ denoting the properly normalized wavefunction. Expanding the commutator, we obtain
\be
\label{eq:Green}
- i \int^{t} \rd t' \;  \langle  [  \phi_3(t) ,  \int \rd^3 x  \; a^4 \; \phi\;{\cal O}^{(2)}(t') ] \rangle = - i \int^{t} \rd t' \; a^4 \left[ \phi_{\rm cl} (t') \phi_{\rm cl}^* (t) - \phi_{\rm cl} (t) \phi_{\rm cl}^* (t') \right] {\cal O}^{(2)}(t') \;.
\ee
Inside the brackets we recognize the Green's function, so that this expression gives exactly the classical non-linear evolution of the field $\phi_3$. In the late time limit this Green's function has two contributions, of the form
\be
t^{\Delta_-} t'^{\Delta_+}~; \qquad~t^{\Delta_+} t'^{\Delta_-} \;.
\ee
The first contribution depends on $t$ as the growing mode solution of the free theory.\footnote{In what follows we assume that $m^2/H^2  < 9/4$ so that $\Delta_\pm$ are real.} Therefore, this contribution will give the time dependence which follows from the assignment of conformal weights provided that the integral in \eqref{eq:Green}, namely 
\be
\int^{t} \rd t' \frac{1}{t'^4} t'^{\Delta_+} {\cal O}^{(2)}(t')\,,
\ee
converges. This, as we discussed, means physically that the classical interaction is dominated by early and not late times. Notice that for $m^2 = 0$ and ${\cal O}^{(2)}(t') =$ const, we have a logarithmic divergence. A positive $m^2$ or an interaction containing derivatives makes the integral converge. What about the second contribution ($\sim t^{\Delta_+} t'^{\Delta_-}$) in the Green's function? This is always suppressed compared with the one we discussed as $|t'| < |t|$, and they become comparable at late times $t' \sim t$. Again, if the above integral converges, this late time limit is irrelevant, and the second contribution to the Green's function is negligible.

\subsection{Off-shell ambiguity}
\label{offshell}

As we have already pointed out, the momenta on the right-hand side of the consistency relation do not form a closed polygon. In other words, the correlation function has to be evaluated slightly ``off-shell". However, 
when calculating the $N$-point function, there are many different ways one can go off-shell.\footnote{We thank Alberto Nicolis for bringing this ambiguity to our attention and for discussions on the matter.} To see this, let us write a generic $N$-point function in the following form
\be
\langle \phi_{\vec k_1}\ldots \phi_{\vec k_N} \rangle = (2\pi)^3\delta(\vec k_1+ \ldots + \vec k_N)\mathcal A' \;.
\ee
Adding a term that is zero on-shell to the amplitude $\mathcal A'$ does not change the result. For an arbitrary function $\vec F$, we can write 
\be
\langle \phi_{\vec k_1}\ldots \phi_{\vec k_N} \rangle = (2\pi)^3\delta^3(\vec P)\left(\mathcal A' + \vec P\cdot \vec F(\vec k_1,\ldots, \vec k_N) \right) \;,
\ee
where, as before, $\vec P = \sum_i \vec k_i$. However, once we go off-shell, it is far from clear that the right-hand side of the consistency relation gives the same result independently on the choice of the function $\vec F$. This has to be the case for the consistency relation to make sense, because the left-hand side is unambiguous. Below we will prove that there is in fact no ambiguity in the consistency relation, and we will show how it can be rewritten in a form where all the correlation functions are calculated ``on-shell".

The simplest check is to explicitly compute the action of the differential operator in the consistency relation on an expression that is zero on-shell and that can be generically written in the form $\vec P \cdot \vec F(\vec k_i)$. A straightforward calculation gives
\begin{align}
& \left(1 + \frac{1}{N} q_i \sum_a \partial_{k_{ai}} + \frac{1}{6N}\vec q^{\;2} \sum_a \partial_{k_a}^2 \) \vec P \cdot \vec F  \nonumber \\
&\quad  = (\vec P + \vec q)\cdot \vec F + \frac{1}{N} \frac{q^2}{3}\sum_a \partial_{k_{ai}} F_i + \frac{1}{N} q_i P_l \sum_a \partial_{k_{ai}} F_l  + \frac{1}{6N}q^2P_i\sum_a \partial^2_{k_a} F_i \;.
\end{align}
Using the fact that $\vec P + \vec q = 0$, we see that the first term on the right-hand side is zero, while the second and third cancel upon averaging. The last term is of order $\mathcal O(q^3)$ and can be neglected. Indeed, as we expected, adding  to the amplitude of an $N$-point function a term that vanishes on-shell does not change the result on the right-hand side of the consistency relation (up to corrections of order $\mathcal O (q^3)$). This proves that there is no ambiguity, and that different choices of going slightly off-shell give the same contribution in the end.

We are now going to give a slightly more general proof of the same statement, which will make more explicit the constraints that different broken generators impose on the relation among correlation functions. The aim is to start from the consistency relation written in form \eqref{resumed1} and rewrite it as a set of different relations that constrain different powers of $\vec q$. If we understand expression \eqref{resumed1} as a Taylor expansion of the $(N+1)$-point function, then this procedure corresponds to isolating the coefficients in this expansion. Since we have constraints on momenta from the delta function on both sides, and we are computing one of the correlation functions slightly off-shell, this procedure is far from trivial. As a first step, let us rewrite the consistency relation (suppressing for simplicity the corresponding time dependence of the modes that is the same as before) in the following form
\be
\frac{1}{P_\pi(q)}\langle \pi_{\vec q}\phi_{\vec k_1}\ldots \phi_{\vec k_N} \rangle'_{q\rightarrow 0} \overset{\rm avg}{=} -t\sum_a \(1 + q_i\partial_{k_{ai}}  + \frac{1}{6}\vec q^{\;2}\partial_{k_a}^2 \) \partial_{t_a} \langle \phi_{\vec k_1} \ldots \phi_{\vec k_N} \rangle' \;.
\ee
At zeroth order in $\vec q$ this relation trivially becomes
\be
\frac{1}{P_\pi(q)}\langle \pi_{\vec q}\phi_{\vec k_1}\ldots \phi_{\vec k_n} \rangle'_{q\rightarrow 0} = -t\sum_a \partial_{t_a} \langle \phi_{\vec k_1} \ldots \phi_{\vec k_n} \rangle' \;,
\label{wardq0}
\ee
where both sides are calculated on-shell. As expected, this expression is the same as \eqref{dilationscons}. The right-hand side can be understood as a zeroth order in a Taylor expansion and, obviously, in this case there is no ambiguity in the consistency relation. To find the next coefficient in the Taylor expansion of the left-hand side, we must differentiate with respect to $\vec q$ and set $\vec q =0$.
\be
\partial_{q_i}\( \frac{1}{P_\pi(q)}\langle \pi_{\vec q}\phi_{\vec k_1}\ldots \phi_{\vec k_N} \rangle'_{q\rightarrow 0} \)\bigg|_{q=0} = -t\sum_{a=1}^N (\partial_{k_{ai}} + \partial_{q_i}) \partial_{t_a} \langle \phi_{\vec k_1} \ldots \phi_{\vec k_N} \rangle' \bigg|_{q=0}\;,
\ee
There are two things to be noticed. Firstly, using the on-shell condition $\vec k_1 + \ldots \vec k_N + \vec q = 0$ we can write $\partial_{q_i} = - \partial_{k_{Ni}}$. Secondly, since we set $\vec q$ to zero, the prime on the correlation function on the right-hand side now means that this correlator is calculated on-shell. The previous expression becomes
\be
\partial_{q_i}\( \frac{1}{P_\pi(q)}\langle \pi_{\vec q}\phi_{\vec k_1}\ldots \phi_{\vec k_N} \rangle'_{q\rightarrow 0} \)\bigg|_{q=0} = -t\sum_{a=1}^{N-1} (\partial_{k_{ai}} - \partial_{k_{Ni}}) \partial_{t_a} \langle \phi_{\vec k_1} \ldots \phi_{\vec k_N} \rangle' \;.
\ee
The difference of the derivatives in the brackets can be rewritten in a form of a total derivative
\be
\partial_{k_{ai}} - \partial_{k_{Ni}} = \partial_{k_{ai}} + \frac{\rmd k_{Nj}}{\rmd k_{ai}} \partial_{k_{Nj}} = \frac{\rmd}{\rmd k_{ai}} \;.
\ee
Since the correlator is on-shell and any dependence on $\vec k_N$ can be removed, we can trivially include into the sum a total derivative with respect to $\vec k_N$. The final expression that we obtain is
\be
\partial_{q_i}\( \frac{1}{P_\pi(q)}\langle \pi_{\vec q}\phi_{\vec k_1}\ldots \phi_{\vec k_N} \rangle'_{q\rightarrow 0} \)\bigg|_{q=0} = -t\sum_{a=1}^{N} \frac{\rmd}{\rmd k_{ai}} \partial_{t_a} \langle \phi_{\vec k_1} \ldots \phi_{\vec k_N} \rangle' \;.
\label{wardq1}
\ee
The last relation represents the consistency relation for broken boosts, and it corresponds to the terms in \eqref{resumed2} that are first order in $\vec q$. The most important point to stress is that, in this form, the correlation functions on both sides are calculated on-shell, and there is therefore no ambiguity in the final result. Finally, we can follow the same procedure to isolate the terms of order $q^2$. 
The resulting expression is
\be
\partial_{q}^2\( \frac{1}{P_\pi(q)}\langle \pi_{\vec q}\phi_{\vec k_1}\ldots \phi_{\vec k_N} \rangle'_{q\rightarrow 0} \)\bigg|_{q=0} = -t\sum_{a=1}^{N} \frac{\rmd^2}{\rmd k_{a}^2} \partial_{t_a} \langle \phi_{\vec k_1} \ldots \phi_{\vec k_N} \rangle' \;.
\label{wardq2}
\ee

Equations \eqref{wardq0}, \eqref{wardq1} and \eqref{wardq2} are equivalent to the consistency relation \eqref{resumed2}. As we have already pointed out, in this form all correlation functions are computed on-shell. This is another proof that in the consistency relation \eqref{resumed2} there is no ambiguity in off-shell prescription on the right side of the equation. These derivative identities are in fact the Ward identities corresponding to the non-linearly realized time translations, boosts and special conformal transformation, respectively. A formal derivation of these relations using the machinery of \cite{LamKurt} will appear in future work~\cite{ward2}.

To perform explicit checks using the derivative form of the consistency relation, we work on-shell on both sides at all times. Therefore, on the left-hand side, we express one of the momenta, say $k_N$, as a sum of the other momenta. We then take the squeezed limit, obtaining a  function of $q$, the small momentum. In order to check the various relations, we can then take derivatives of this left hand side with respect to $q$ and then set $q=0$. On the right-hand side, we must also work on-shell. This means that we also write the $k_N$ momentum in terms of the other $N-1$ momenta (not including $q$).

Schematically, the procedure is as follows: consider checking the consistency relation
\be
\partial_q^2\left(\frac{1}{P_\pi(q)}\langle\pi_{\vec q}\phi_{\vec k_1}\ldots\phi_{\vec k_N}\rangle'\right) = -\frac{1}{N}t\sum_{a=1}^N\frac{\rd}{\rd k_a^2}\frac{\rd}{\rd t}\langle\phi_{\vec k_1}\ldots\phi_{\vec k_N}\rangle'~.
\ee
We rewrite the left hand side so that it is a function of $N$ different momenta, that is we take $\vec k_N = -\sum \vec k_a-\vec q$. We then take the squeezed limit $q\to0$ and differentiate with respect to $q$. On the right hand side, we write $\vec k_N = -\sum \vec k_a$. This means that we actually only have to take $N-1$ derivatives on the right hand side.

For illustrative purposes, we provide an explicit check of the consistency relation in differential form. Consider the soft limit of the three-point function involving only $\pi$ fields, $\langle\pi^3\rangle$. The three and two-point correlation functions are given by
\be
\langle\pi_{\vec q}\pi_{\vec k_1}\pi_{\vec k_2}\rangle' = \frac{81H^4}{4M_\pi^4}\frac{\left(q^5+k_1^5+k_2^5\right)}{q^5k_1^5k_2^5 t^4}~;\qquad \langle\pi_{\vec{k}_1}\pi_{\vec{k}_2}\rangle' = \frac{9H^2}{2M_\pi^2}\frac{1}{k_1^5t^2}\,.
\ee
We take the limit $q\to 0$ to obtain the squeezed limit of the three-point function
\be
\frac{1}{P_\pi(q)}\langle\pi_{\vec q}\pi_{\vec k_1}\pi_{\vec k_2}\rangle' = \frac{9H^2}{2M_\pi^2t^2}\left[\frac{2}{k_1^5}+\frac{5(\vec q\cdot\vec k_1)}{k_1^7}+\frac{5\left(7(\vec q\cdot\vec k_1)^2-q^2k_1^2\right)}{2k_1^9}
\right]+{\cal O}(q^3)~.
\ee
From this, we immediately read off:
\bea
\nonumber
\frac{1}{P_\pi(q)}\langle\pi_{\vec q}\pi_{\vec k_1}\pi_{\vec k_2}\rangle'_{q\to0} &=&  \frac{9H^2}{M_\pi^2}\frac{1}{k_1^5t^2} =  -t\frac{\rd}{\rd t}\langle\pi_{\vec{k}_1}\pi_{\vec{k}_2}\rangle' \;; \\
\frac{\partial}{\partial q^i}\left(\frac{1}{P_\pi(q)}\langle\pi_{\vec q}\pi_{\vec k_1}\pi_{\vec k_2}\rangle' \right) \bigg|_{q=0} &=&  \frac{45H^2}{M_\pi^2}\frac{k_1^i}{k_1^7t^2} = -\frac{1}{2}t\frac{\rd}{\rd k_1^i}\frac{\rd}{\rd t}\langle\pi_{\vec{k}_1}\pi_{\vec{k}_2}\rangle'   \;; \\
\partial_q^2 \left(\frac{1}{P_\pi(q)}\langle\pi_{\vec q}\pi_{\vec k_1}\pi_{\vec k_2}\rangle' \right) \bigg|_{q=0} &=& \frac{90H^2}{M_\pi^2}\frac{1}{k_1^7t^2} =-\frac{1}{2}t\frac{\rd^2}{\rd k_1^2}\frac{\rd}{\rd t} \langle\pi_{\vec{k}_1}\pi_{\vec{k}_2}\rangle'\,,\nonumber
\eea
where the last step in each equation follows from differentiating $ \langle\pi_{\vec{k}_1}\pi_{\vec{k}_2}\rangle'$. Thus the derivative form of the consistency relation checks out at each order. Note that when checking the identities in Ward form, it is {\it not} necessary to take an angular average in the ${\cal O}(q^2)$ terms. The Laplacian, $\partial_q^2$, appearing on the left hand side automatically picks out the trace part of the order $q^2$ terms of the soft mode.

\section{Conclusions}
\label{conclu}

Symmetry considerations allow us to make powerful, model-independent statements. In this paper, we have considered how non-linearly realized symmetries in the conformal mechanism enforce various relations among the correlation functions of the theory. In doing so, we have uncovered some model-independent and robust predictions of the mechanism.  It is worth summarizing them:
\begin{itemize}
\item Absence of detectable gravitational waves.
\item Model-dependent local non-Gaussianity from the conversion mechanism.
\item Anisotropy of the power spectrum, see~\eqref{totalanisotropy}, \cite{Libanov:2010nk, Libanov:2010ci, Libanov:2011bk}.
\item 4-point function in the soft internal limit due to tree-level $\pi$ exchange,~\eqref{Ruby},  \cite{Libanov:2010nk, Libanov:2010ci, Libanov:2011bk}. This is relevant on large scales.
\item 4-point function in the soft internal limit due to one-loop $\pi$ exchange,~\eqref{oneloop}. This dominates for sufficiently small internal momentum, and it shows up as stochastic bias and in the power spectrum of $\mu$ distortion. 
\end{itemize}
It will be interesting to study which of the last three signatures is the most constraining, as they all depend on the same parameters. From our rough estimates and the conclusions of \cite{Ramazanov:2012za}, it seems that the one-loop four-point function is the most constraining, as it gives ${\cal} P_\pi \lesssim 1$. However, all the experimental constraints are not optimized for the particular shapes of the four-point functions we got, so that numbers could change sizably when a dedicated analysis is implemented.

It will also be interesting to further elucidate the properties of correlation functions when more than one external leg is taken to be soft. These higher-soft limits can serve to probe the underlying broken symmetry algebra, and could provide another set of consistency relations, not just for the conformal mechanism, but also in the case of inflation.

Another interesting direction to pursue is to derive the consistency relations for the DBI version of the conformal mechanism studied in \cite{Hinterbichler:2012fr, Hinterbichler:2012yn}. In this scenario, the symmetry breaking pattern is the same, but the non-linearly realized conformal symmetries act in a different way. It would be interesting to understand if the consistency relations are the same in this case, and whether the generic predictions of the conformal mechanism, such as statistical anisotropy, go through in this case.\\
\\
{\bf Acknowledgements:} We thank Christian Byrnes, Kurt Hinterbichler, Lam Hui, Diana L\'opez Nacir, Alberto Nicolis, and Enrico Trincherini for helpful discussions. We are especially grateful to  Valery Rubakov for detailed feedback on a previous draft. M.~S.~thanks the warm hospitality of the University of Pennsylvania where a part of this project was finished. This work is supported in part by
the NASA ATP grant NNX11AI95G (A.J. and J.K.), the Alfred P. Sloan Foundation and NSF CAREER Award PHY-1145525 (J.K.)

\appendix
\section{Conformal transformations on correlation functions}
In this Appendix we derive the action in Fourier space of the linearly-realized dilation and spatial special conformal transformations on correlation functions.

\label{correlatoroperators}
\subsection{Dilation}
\label{dilcorrelator}
We will work in an arbitrary number of dimensions, $d$. The dilation operator acts linearly on fields in position space as
\be
\delta_D\phi = (\Delta-x^A\partial_A)\phi~.
\ee
We note that the field $\phi$ can be written in Fourier space using
\be
\phi(x)=\int\rd^dk e^{ik\cdot x}\phi_k~,
\ee
we may therefore write
\be
\delta_D\phi = \int\rd^dk\phi_k(\Delta+ \vec k\cdot \vec\partial_k)e^{ik\cdot x}~.
\ee
Now, we can integrate by parts to obtain two terms
\be
\delta_D\phi = \int\rd^dk e^{ik\cdot x}\left(\Delta-d-\vec k\cdot\vec\partial_k\right)\phi_k~.
\ee
From this, we deduce the Fourier space transformation rule
\be
\delta_D\phi_k = -\left((d-\Delta)+\vec k\cdot\vec\partial_k\right)\phi_k
\ee
Now, we want to obtain the action of dilation on a correlation function. A correlation function has two parts, the amplitude and the delta function, schematically it is of the form
\be
\delta_D{\cal A}=\delta_D\left(\delta^3(\vec P){\cal A'}\right)~,
\ee
where the prime indicates removal of the delta function and $\vec P$ is the sum of the momenta $\vec P = \sum \vec k$. We may then write
\begin{align*}
\delta_D\left(\delta^3(\vec P){\cal A'}\right) &= -\sum_{a=1}^N\left((d-\Delta_a)+\vec k_a\cdot\vec\partial_{k_a}\right)\left[\delta^3(\vec P){\cal A}'\right] \\
&= -{\cal A}'\vec P\cdot\vec\partial_P\delta^3(\vec P)-\sum_{a=1}^N\delta^3(\vec P)\left((d-\Delta_a)+\vec k_a\cdot\vec\partial_{k_a}\right){\cal A}'~.
\end{align*}
The term outside the sum may be integrated by parts to obtain a factor of $d$. The term where the derivative $\vec \partial_P$ hits ${\cal A}'$ vanishes because it multiplies $\vec P\delta^3(\vec P) = 0$. We then have
\be
\delta_D\left(\delta^3(\vec P){\cal A'}\right) = \delta^3(\vec P)\left[d-\sum_{a=1}^N\left((d-\Delta_a)+\vec k_a\cdot\vec\partial_{k_a}\right){\cal A}'\right]~.
\ee
From this, we deduce that the dilation operator acts on the amplitude without the delta function as
\be
\delta_D{\cal A}' = \left[-d(N-1)+\sum_{a=1}^N\left(\Delta_a-\vec k_a\cdot\vec\partial_{k_a}\right)\right]{\cal A}'
\ee

\subsection{Special conformal transformations}
\label{confcorrelationaction}
Special conformal transformations act in real space as
\be
\delta_{K^A} = (2\Delta x^A-2x_Ax^B\partial_B+x^2\partial_A)\phi~.
\ee
Following the same steps as above, we may write this in Fourier space acting on the primed correlator as
\be
\delta_{K^A}{\cal A}' = i\sum_{a=1}^N\left(2(\Delta_a-d)\partial_{k_a^A}+k^A_a\vec\partial_{k_a}^2-2\vec k_a\cdot\vec\partial_{k_a}\partial_{k_a^A}\right){\cal A}'
\ee

\section{Construction of actions for spontaneously broken conformal symmetry}
\label{constructionofactions}

In this section we review the construction of actions for the spontaneous breaking of the conformal group down to its de Sitter subgroup \cite{Hinterbichler:2012mv}. Equivalently, this construction gives actions for fields on de Sitter space that non-linearly realize conformal symmetry. The basic idea is fairly simple, in order to linearly realize the isometries of de Sitter, we consider a de Sitter metric $g_{\mu\nu}^{\rm dS}$. A scalar field action constructed using this metric will be invariant under the de Sitter group. In order to non-linearly realize conformal symmetry, we add the conformal mode to this metric and consider
\be
g_{\mu\nu} = e^{2\pi}g_{\mu\nu}^{\rm dS}~.
\label{confmetricds}
\ee
\subsection{Action for the Goldstone}
In order to construct the action for the field $\pi$ --- the Goldstone mode of broken conformal symmetry --- we construct curvature invariants from the metric $g_{\mu\nu}$. Since the metric is conformally flat, all of the information is contained in the Ricci tensor
\be
R_{\mu\nu} = 3H^2g_{\mu\nu}^{\rm dS} -2\nabla_\mu\nabla_\nu\pi-g_{\mu\nu}^{\rm dS}\square\pi+2\partial_\mu\pi\partial_\nu\pi-2g_{\mu\nu}^{\rm dS}(\partial\pi)^2~.
\ee
Then, in order to construct actions for $\pi$, we merely form any diffeomorphism scalar using the conformal metric, its covariant derivative and the Ricci tensor. The lowest order lagrangian is given by the invariant volume and the kinetic term is given by the Ricci scalar
\be
{\cal L}_\pi \sim \sqrt{-g}\left(R-6H^2\right) = \sqrt{-g_{\rm dS}}\left(\frac{1}{2}e^{2\pi}(\partial\pi)^2+\frac{1}{2}e^{2\pi}\square\pi-H^2e^{2\pi}+\frac{H^2}{2}e^{4\pi}\right)~,
\ee
Where we have added the cosmological term so that there will be no tadpole about $\pi=0$. Expanding this lagrangian out to cubic order and integrating by parts reproduces \eqref{picubicaction} in the text. In \cite{Hinterbichler:2012mv}, it was shown that this construction is equivalent to the coset construction of Callan, Coleman, Wess, Zumino and Volkov \cite{Coleman:1969sm,Callan:1969sn, volkov}. This construction misses a four-derivative Wess--Zumino term (see \cite{Goon:2012dy} for details).

\subsection{Coupling matter fields}
It is also straightforward to couple matter fields to the Goldstone field $\pi$; one thing we can do is use the covariant derivative of the conformal metric \eqref{confmetricds} to construct actions for matter fields $\phi$, contracting indices with $g_{\mu\nu}$. Additionally, we are allowed to multiply curvature invariants constructed from $R_{\mu\nu}$ by arbitrary polynomial functions of $\phi$.\footnote{Note that since the WZ term shifts by a total derivative, we are {\it not} allowed to multiply it by a function.} The two-derivative lagrangian for a scalar field is then of the schematic form
\be
{\cal L}_{\rm \phi} \sim \sqrt{-g}\left(-\frac{1}{2}(\partial\phi)^2 -V(\phi)+f(\phi)R\right) = \sqrt{-g_{\rm dS}}\left(-\frac{1}{2}e^{2\pi}(\partial\phi)^2-e^{4\pi}V(\phi)+e^{4\pi}f(\phi)R\right)~.
\ee
Since our aim is to merely check the consistency relations in a variety of examples, we consider the case where $V(\phi) = \frac{m_\phi^2}{2}\phi^2+\lambda\phi^4$ and $f(\phi) = 0$. The lagrangian then takes the form
\be
{\cal L}_{\phi} \sim  \sqrt{-g_{\rm dS}}\left(-\frac{1}{2}e^{2\pi}(\partial\phi)^2-\frac{m_\phi^2}{2}e^{4\pi}\phi^2-\lambda e^{4\pi}\phi^3\right)~.
\label{spectatoraction}
\ee
Expanding about $\phi = \pi = 0$ to quartic order yields the action
\be
S_{\rm \phi} = M_\phi^2\int\rd^4x\sqrt{-g_{\rm dS}}\left(-\frac{1}{2}(\partial\phi)^2-\frac{m_\phi^2}{2}\phi^2-2m_\phi^2\pi\phi^2-\pi(\partial\phi)^2-\lambda\phi^3-4\lambda\pi\phi^3\right)~.
\label{generalspectatoraction}
\ee

\subsection{Transformation of $\pi$}
It is also quite straightforward to work out the way the non-linearly realized conformal symmetries act on $\pi$. Under an infinitesimal diffeomorphism, the metric changes by the Lie derivative
\be
\delta g_{\mu\nu} = -\pounds_{\xi} ~g_{\mu\nu} = -g_{\rho\nu}\nabla_{\mu}\xi^{\rho}-g_{\mu\rho}\nabla_{\nu}\xi^{\rho}~.
\ee
We assume that the background metric $g^{\rm dS}_{\mu\nu}$ remains fixed (this restricts us to isometries of de Sitter plus conformal transformations), so we have
\be
2\delta\pi g_{\mu\nu} = -g_{\rho\nu}\nabla_{\mu}\xi^{\rho}-g_{\mu\rho}\nabla_{\nu}\xi^{\rho}~,
\ee
tracing over both sides gives $\delta\pi = -\frac{1}{4}\nabla_\rho\xi^\rho$. This is the divergence of a vector, so we may write
\be
\delta \pi  = -\frac{1}{4\sqrt{-g}}\partial_\rho\left(\sqrt{-g}\xi^\rho\right) = -\xi^\rho\partial_\rho\pi - \frac{1}{4}\nabla^{\rm dS}_\rho\xi^\rho~.
\ee
So we have the transformation rule for $\pi$,
\be
\delta\pi = -\xi^\rho\partial_\rho\pi - \frac{1}{4}\nabla^{\rm dS}_\rho\xi^\rho~.
\ee
From this transformation rule, it is clear that that $\pi$ will transform linearly under isometries of the dS metric $(\nabla^{\rm dS}_{\rho}\xi^\rho=0)$ and will transform in a nonlinear fashion under broken transformations. To make this explicit, we must make a choice of de Sitter slicing. Choosing the planar inflationary slicing:
\be
g_{\mu\nu}^{\rm dS} = \frac{1}{H^2t^2}t_{\mu\nu}~,
\ee
we find
\be
\delta \pi = -\xi^{\rho}\partial_\rho\pi - \frac{1}{4}\partial_{\mu}\xi^\mu+\frac{1}{t}\xi^0~.
\ee
From this, it is straightforward to derive the symmetries \eqref{pitrans}.

\section{Correlation functions}
\label{correlationfncomputation}

Here we collect some results for correlation functions involving spectator fields coupled to the Goldstone field $\pi$.

\subsection{Mode functions for massive fields}

In this Appendix, we derive the expression for the mode functions of a massive scalar field on de Sitter space in terms of Hankel functions. This expressions are needed to compute the correlation functions we need to check the consistency relations.
Consider the general quadratic action for a massive scalar
\be
S_{2, \phi} = M_\phi^2\int \rd^4 x\sqrt{-g}\left(-\frac{1}{2}(\partial\phi)^2-\frac{m_\phi^2}{2}\phi^2\right)~.
\ee
Where $m_\phi^2$ is an arbitrary mass. The equation of motion following from this action is
\be
\square\phi+\frac{2}{t}\dot\phi-\frac{m_\phi^2}{H^2t^2}\phi = 0~.
\ee
We define the canonically-normalized variable
\be
v = \frac{M_\phi}{H(-t)}\phi~,
\ee
whose mode functions satisfy
\be
v_k''+\left[k^2-\left(2-\frac{m_\phi^2}{H^2}\right)\frac{1}{t^2}\right]v_k=0~.
\ee
Defining $x\equiv -kt$ and $\nu \equiv \sqrt{\frac{9}{4}-\frac{m_\phi^2}{H^2}}$, after changing variables to $f_k \equiv v_k/\sqrt{x}$ this can be cast as Bessel's equation
\be
x^2\frac{\rd^2 f_k}{\rd x^2} +x\frac{\rd f_k}{\rd x}+(x^2-\nu^2)f_k=0~,
\ee
which is well-known to be solved by (we choose Hankel functions as our basis)
\be
f_k(x) = c_1(k)H^{(1)}_\nu(x)+c_2(k)H_\nu^{(2)}(x)~.
\label{fmodefunctions}
\ee
We fix the coefficients by demanding that in the far past $(-kt\to\infty)$, the mode functions of the canonically normalized variable, $v_k$, have their Minkowski space form. This is the so-called adiabatic vacuum (Bunch--Davies) choice. That is, we demand
\be
v_k(t) \underset{-kt\to\infty}{\longrightarrow} \frac{1}{\sqrt{2 k}}e^{-ikt}
\ee
Then, using the asymptotic expansion for the Hankel functions as $-kt \to\infty$
\begin{align*}
H^{(1)}_\nu(-kt) &\sim -e^{\frac{i\pi}{2}\left(\frac{3}{2}-\nu\right)}\sqrt{\frac{2}{\pi}}\frac{1}{\sqrt{-kt}}e^{-ikt}\\
H^{(2)}_\nu(-kt) &\sim e^{\frac{i\pi}{2}\left(\frac{1}{2}+\nu\right)}\sqrt{\frac{2}{\pi}}\frac{1}{\sqrt{-kt}}e^{ikt}
\end{align*}
This implies that we need to take $c_1(k) = -e^{-\frac{i\pi}{2}\left(\frac{3}{2}-\nu\right)}\sqrt{\frac{\pi}{4}}\frac{1}{\sqrt{k}}$ and $c_2(k) = 0$ in (\ref{fmodefunctions}). This leads to the expression for the $\phi_k$ mode functions
\be
\phi_k(t) =  -e^{-\frac{i\pi}{2}\left(\frac{3}{2}-\nu\right)}\sqrt{\frac{\pi}{4}}\frac{H(-t)^{3/2}}{M_\phi}H_\nu^{(1)}(-kt)~~~~~~{\rm with}~~~\nu = \sqrt{\frac{9}{4}-\frac{m_\phi^2}{H^2}}~,
\ee
where $H_\nu^{(1)}(-kt)$ is a Hankel function of the first kind. Note that for $m^2 > \frac{9H^2}{4}$ the solution is a Hankel function of imaginary order.

\subsection{In-in integrals}

In order to compute correlation functions, we employ the Schwinger--Keldysh or in-in formalism (see \cite{Maldacena:2002vr, Weinberg:2005vy} for an exposition). In this formalism, rather than computing in-out S-matrix elements, we compute correlation functions sandwiched between the same vaccum. The correlation function for an operator, ${\cal O}(t)$ is given by \cite{Maldacena:2002vr, Weinberg:2005vy}
\be
\langle{\cal O}(t)\rangle = \langle 0\rvert \bar Te^{i\int_{t_0}^t\rd t' H_{\rm int}(t')}{\cal O}(t) T e^{-i\int_{t_0}^t\rd t' H_{\rm int}(t')}\lvert 0\rangle~.
\ee
Here $H_{\rm int}$ is the interaction Hamiltonian, $T$ denotes time-ordering while $\bar T$ denotes anti-time-ordering and $t_0$ is an early time. Generally we will only work to leading order (tree-level) where the correlation function is given by
\be
\langle{\cal O}(t)\rangle = -i\int_{-\infty}^{t}\rd t'\left\langle0\left\rvert\left[{\cal O}(t), H_{\rm int}(t')\right]\right\lvert0\right\rangle~.
\label{inintreelevel}
\ee

 \subsection{Correlation functions of $\pi$}
Here we compute the two and three-point correlators for the Goldstone field $\pi$. We consider the action \eqref{picubicaction}. The quadratic equations of motion lead to the following mode function for the field $\pi$
\be
\pi_{k}(t) = -i\frac{H(-t)^{3/2}}{M_\pi}\sqrt\frac{\pi}{4}H^{(1)}_{5/2}(-kt) = \frac{-3H}{\sqrt{2 k^5}(-t)M_\pi}\left(1+ikt-\frac{k^2t^2}{3}\right)e^{-ikt}~.
\label{pimodefunction}
\ee
From this the two-point function can straightforwardly be computed:
\be
\label{pipi}
P_\pi(k) \equiv \langle\pi_{\vec k}\pi_{-\vec k}\rangle' = \frac{9H^2}{2M_\pi^2k^5 t^2}\left(1+\frac{k^2t^2}{3}+\frac{k^4t^4}{9}\right)~.
\ee
Note that this field has an extremely red spectrum, peaking strongly as $k\to 0$.

From the action (\ref{picubicaction}), we can  also compute the three-point function, $\langle\pi^3\rangle$. The interaction Hamiltonian, $H_{\rm int}$, at this order is minus the lagrangian
\be
H_{\rm int} = -\int\rmd^3x {\cal L}_{\rm int} = M_\pi^2\int\rmd^3x \left[\frac{1}{H^2t^2}\pi(\partial\pi)^2-\frac{4}{H^2t^4}\pi^3\right]~.
\ee
Applying the formula \eqref{inintreelevel}, we obtain (at late times)
\be
\langle\pi_{\vec k_1}\pi_{\vec k_2}\pi_{\vec k_3}\rangle' = \frac{81H^4}{4M_\pi^4}\frac{\left(k_1^5+k_2^5+k_3^5\right)}{k_1^5k_2^5k_3^5 t^4}~.
\label{pi3ptfn}
\ee

\subsection{Massive spectator field, $\Delta = 1$}
The simplest case of a spectator field coupled to $\pi$ is a massive field with $m_\phi^2 \equiv m_\vp^2 = 2H^2$, corresponding to $3d$ conformal weight $\Delta =1$. We take the action \eqref{generalspectatoraction} with this choice of mass:
\be
S_\vp = M_\vp^2\int\rd^4 x\sqrt{-g}\left(-\frac{1}{2}(\partial\vp)^2-H^2\vp^2-4H^2\pi\vp^2-\pi(\partial\vp)^2-\lambda\vp^3-4\lambda\pi\vp^3\right)~,
\ee
The mode functions for the field are given by
\be
\vp_k(t) = \frac{iH(-t)}{\sqrt{2k}M_\vp}e^{ikt}~,
\ee
which leads to the two-point function for the spectator
\be
\label{phiphi}
P_\vp(k) \equiv \langle\vp_{\vec k}\vp_{-\vec k}\rangle' = \frac{H^2}{2M_\vp^2}\frac{t^2}{k}~.
\ee
We can also compute various higher-point correlation functions involving this spectator. The simplest is the three-point function involving only $\vp$, the tree-level correlation function is given by
\be
\langle\vp_{\vec k_1}\vp_{\vec k_2}\vp_{\vec k_3}\rangle' =  \frac{3\pi H^2\lambda}{4M_\vp^4}\frac{t^3}{k_1k_2k_3}~.
\label{vpvpvp}
\ee
Additionally, we can compute the $\langle\pi\vp\vp\rangle$ three-point function for these fields. There are two contributions to the correlation function, one from each of the $\pi\vp\vp$ vertex and the $\pi(\partial\vp)^2$ vertex; the final result is given by
\be
\langle\pi_{\vec q}\vp_{\vec k_2}\vp_{\vec k_2}\rangle' = -\frac{9H^4}{4M_\vp^2M_\pi^2}\frac{1}{q^5k_1k_2}(k_1+k_2)~.
\label{piphiphiappendix}
\ee
This correlation function is invariant under ($4d$) dilations and under $\delta_{K^i}$ with $\Delta_a = \{-1,1,1\}$, agreeing with our general arguments for when conformal weights may be consistently defined, in spite of the fact that this correlation function does not scale in the na\"ive way with time.

Finally, we compute a four-point function, involving three $\vp$ fields and one Goldstone; this computation is slightly more involved. There are two contributions to this four-point function, one coming from a contact diagram involving the $\pi\vp^3$ vertex and one coming from an exchange diagram at second order in the vertices involving a single $\pi$ and two $\vp$'s. The interaction Hamiltonian is given by\footnote{Note that at this order, we must be careful in deriving the interaction Hamiltonian, in this case it is still minus the interaction lagrangian, but in general this will not be true at quartic order.}
\begin{align}
\nonumber
H_{\rm int}^{(3)} &= M_\vp^2\int\rd^3x \left(-\frac{1}{H^2t^2}\pi\dot\vp^2+\frac{4}{H^2t^4}\pi\vp^2+\frac{\lambda}{H^4t^4}\vp^3\right)\\
H_{\rm int}^{(4)} &= M_\vp^2\int\rd^3x \left(\frac{4\lambda}{H^4t^4}\pi\vp^3\right)~.
\end{align}
The correlation function is then a sum of three terms
\begin{align}
\nonumber
\langle\pi_{\vec q}\vp_{\vec k_1}\vp_{\vec k_2}\vp_{\vec k_3}\rangle &= -i\int_{-\infty}^{t}\rd t'\langle0\rvert[\pi_{\vec q}\vp_{\vec k_1}\vp_{\vec k_2}\vp_{\vec k_3}(t), H^{(4)}_{\rm int}(t')]\lvert0\rangle\\
&~~+ \int_{-\infty}^t\rd t'\int_{-\infty}^t\rd t''\langle 0\rvert H_{\rm int}^{(3)}(t')\pi_{\vec q}\vp_{\vec k_1}\vp_{\vec k_2}\vp_{\vec k_3}(t)H_{\rm int}^{(3)}(t'')\lvert 0\rangle\\
\nonumber
&~~-2{\rm Re}\left(\int_{-\infty}^t\rd t'\int_{-\infty}^{t'}\rd t''\langle0\rvert\pi_{\vec q}\vp_{\vec k_1}\vp_{\vec k_2}\vp_{\vec k_3}(t)H_{\rm int}^{(3)}(t')H_{\rm int}^{(3)}(t'')\lvert0\rangle\right)~.
\end{align}
When the dust settles, the four-point function is given by
\be
\label{4pfpichichichi}
\langle\pi_{\vec q}\vp_{\vec k_1}\vp_{\vec k_2}\vp_{\vec k_3}\rangle' = -\frac{27\pi H^4\lambda}{8M_\pi^2M_\vp^4}\frac{t}{q^5k_1k_2k_3}\left(\frac{k_1}{\lvert\vec q+\vec k_1\rvert}+\frac{k_2}{\lvert\vec q+\vec k_2\rvert}+\frac{k_3}{\lvert\vec q+\vec k_3\rvert}\right)~.
\ee

\subsection{Massless spectator field, $\Delta =0$}
We now consider a massless spectator field, corresponding to  \eqref{generalspectatoraction} with with $m_\chi^2=\lambda=0$. The cubic action for this field is given by
\be
S_\chi= M_\chi^2\int\rd^4 x\sqrt{- g}\left(-\frac{1}{2}(\partial\chi)^2-\pi(\partial\chi)^2\right)~.
\label{chicubicaction}
\ee
The mode functions for $\chi$ are the well-known result for massless fields
\be
\chi_{\vec k}(t) = \frac{H}{\sqrt{2k^3}M_\chi}\left(1-ikt\right)e^{ikt}~.
\label{chimodefunction}
\ee
Using this, the two point function for a massless field is the standard result
\be
P_\chi(k) \equiv \langle\chi_{\vec k}\chi_{-\vec k}\rangle' = \frac{H^2}{2k^3 M_\chi^2}(1+k^2t^2)~.
\ee
Additionally, we can compute the three-point function $\langle\pi\chi\chi\rangle$ using the standard techniques, summarized above. At late times, we obtain
\begin{align}
\label{3pfmassless}
\langle\pi_{\vec q}\chi_{\vec k_1}\chi_{\vec k_2}\rangle' &= \frac{3\pi H^4}{16M_\pi^2M_\chi^2}\frac{1}{q^5k_1^3k_2^3 t}\bigg(q^4+2q^2(k_1^2+k_2^2)-3(k_1^2-k_2^2)^2\bigg)\\\nonumber
&-\frac{9H^4}{8M_\chi^2M_\pi^2}\frac{1}{q^5k_1^3k_2^3}\bigg(q^2(k_1^3+k_2^3)-(k_1^5+k_2^5)+3(k_1^3k_2^2+k_1^2k_2^3)\bigg)~.
\end{align}
This correlation function is invariant under $\delta_{K^i}$ with $\Delta_a = \{-1,0,0\}$. Additionally, it has the leading scaling behavior with respect to time that is expected.

\end{document}